\documentclass[12pt]{article}
\usepackage{pstricks}
\usepackage{color,xcolor}
\usepackage{amssymb,amsmath,bm,bbm}
\usepackage{epsf}
\usepackage{epsfig}
\usepackage{afterpage}
\usepackage{longtable}
\usepackage{cite}
\usepackage{latexsym, mathrsfs}
\usepackage{graphics}
\usepackage{url}
\usepackage{paralist}
\usepackage{bbold}
\usepackage{cleveref}
\usepackage{slashed}

\usepackage{braket}

\newcommand{\be}{\begin{equation}}
\newcommand{\ee}{\end{equation}}
\newcommand{\bea}{\begin{eqnarray}}
\newcommand{\eea}{\end{eqnarray}}
\newcommand{\nn}{\nonumber}
\newcommand{\dd}{\displaystyle}

\newcommand{\rvecD}{\overrightarrow D}
\newcommand{\lvecD}{\overleftarrow D}
\newcommand{\noi}{\noindent}
\DeclareRobustCommand{\rchi}{{\mathpalette\irchi\relax}}
\newcommand{\irchi}[2]{\raisebox{\depth}{$#1\chi$}}

\newcommand{\tr}{\text{Tr}}

\newcommand{\Dleft}{\overleftarrow{D}}
\newcommand{\Dright}{\overrightarrow{D}}
\newcommand{\lagr}{\mathcal{L}}
\newcommand{\diff}{\text{d}}
\newcommand{\tprod}{\text{T}}

\begin{document}

\begin{flushright} {BARI-TH/22-734}\end{flushright}

\medskip

\begin{center}
{\Large  Semileptonic $B_c$ decays to $P-$wave charmonium\\ \vspace*{0.3cm}
and the nature of  $\chi_{c1}(3872)$ }
\\[1.0 cm]
{ {P.~Colangelo$^a$, F.~De~Fazio$^a$,   F.~Loparco$^a$ , N.~Losacco$^{a,b}$, M.~Novoa-Brunet$^{a}$ }
 \\[0.5 cm]}
{\small 
$^a$Istituto Nazionale di Fisica Nucleare, Sezione di Bari,  Via Orabona 4, I-70126 Bari, Italy \\[0.1 cm]
$^b$Dipartimento Interateneo di Fisica ``M. Merlin'', Universit\`a  e Politecnico di Bari, \\ via Orabona 4, 70126 Bari, Italy
}
\end{center}

\vskip 0.8cm

\begin{abstract}
\noindent
The hadronic form factors of $B_c$ semileptonic decays to the $P-$wave charmonium 4-plet can be expressed near the zero-recoil point in terms of universal functions, performing a  systematic expansion in QCD in the relative velocity of the heavy quarks and in $1/m_Q$. Such functions are independent of the member of the multiplet involved in the transitions. We present the results of an NLO calculation up to $O(1/m_Q^2)$ classifying the universal functions at this order. We work out a set of relations among  the form factors of the same mode and of different modes, which should be reproduced by explicit calculations, reducing the hadronic uncertainty affecting such channels. The approach  is also helpful to  investigate the debated nature of  $\chi_{c1}(3872)$, studying the production in $B_c$ semileptonic decays and comparing it to the modes involving the other $2P$ charmonia.

\end{abstract}

\thispagestyle{empty}


\section{Introduction}
There is a manifold interest in the study of semileptonic decays of heavy hadrons, namely those induced by the $b \to c \ell \bar \nu_\ell$ transition at the quark level. The first one is the possibility to precisely measure fundamental parameters of the Standard Model (SM), in this case the element $|V_{cb}|$ of the Cabibbo-Kobayashi-Maskawa (CKM) mixing matrix. The analyses of processes involving different hadrons, modes and final states (exclusive/inclusive) should provide compatible results:  such a compatibility has not been  achieved, yet, considering the debated $|V_{cb}|$ determinations from inclusive and exclusive $B$ decays \cite{Gambino:2020jvv} which fuel discussions on  possible explanations of the tension within the  SM  \cite{Martinelli:2021myh} or beyond \cite{Colangelo:2016ymy}. Other processes, e.g.  $B_c$ to charmonium, give access to such a fundamental parameter which has tight correlations with  flavour observables \cite{Buras:2022nfn}.

Another important interest relies on the possibility of testing fundamental features of the Standard Model, namely  Lepton Flavour Universality (LFU). Signals of LFU violation have been detected in $B$ decays induced by this transition \cite{HFLAV:2022pwe}. They hint to physics beyond SM   \cite{Barbieri:2021wrc}, the structure of which can be constrained by studying  sets of decay observables 
\cite{Biancofiore:2013ki,Alonso:2016oyd,Colangelo:2018cnj,Jung:2018lfu,Murgui:2019czp,Becirevic:2016hea,Alguero:2020ukk,Cornella:2021sby}. Related LFU violating effects should be observed in 
processes involving different hadrons: the first measurements are available for $B_c \to J/\psi \ell \bar \nu_\ell $  \cite{Aaij:2017tyk},  other $B_c$ modes can be considered for such  investigations.

In addition to the  above motivations concerning the structure of the theory of electroweak interaction and of its extensions, there is interest related to  strong interaction effects. For decays involving hadrons comprising a single heavy quark $Q$, a double  expansion in QCD in powers of $1/m_Q$ and of  $\alpha_s$  provides a powerful method to classify   the hadronic matrix elements, both in the exclusive and  inclusive transitions. This is the basis for an efficient  control of the theoretical uncertainty in the above mentioned  measurements. For mesons comprising two heavy quarks such as $B_c$, the expansion parameter is the relative three-velocity of the heavy quarks, with counting rules  given within  Nonrelativistic QCD (NRQCD).  At the various orders in the expansion  the form factors  governing   $B_c \to J/\psi (\eta_c ) \ell \bar \nu_\ell$  can be given in terms of universal functions, the same for the two final states,  in selected kinematical ranges \cite{Jenkins:1992nb,Colangelo:1999zn,Colangelo:2022lpy}.
Relations can be established among the form factors,
with a reduction of the hadronic uncertainties affecting  the description of the semileptonic channels. It is worth considering such an expansion also for the form factors governing the $B_c$ transitions to the $P$-wave charmonia  involving a spin 4-plet of final states, for which little information is available at present.\footnote{The peculiar role of $B_c$, a weakly decaying meson comprising only heavy quarks, has been recently discussed in \cite{Colangelo:2021dnv,Colangelo:2021myn}.}

There is a further reason of interest.
Semileptonic decays provide us with a probe on the structure of the hadrons involved in the transitions. This is important, for example,  if one considers   the meson $\chi_{c1}(3872)$ (usually denoted as  $X(3872)$). After the observation by the Belle Collaboration \cite{Belle:2003nnu} and the confirmation and  measurements by other collaborations  \cite{CDF:2003cab,D0:2004zmu, BaBar:2004oro,CDF:2005cfq,CDF:2009nxk,LHCb:2011zzp,BESIII:2013fnz,CMS:2013fpt,ATLAS:2016kwu},   such a meson, whose quantum numbers are  $J^{PC}=1^{++}$ \cite{LHCb:2013kgk,LHCb:2015jfc},  is under intense scrutiny due to features hinting a non conventional  charmonium structure. Puzzling properties are the closeness of the mass of  $X(3872)$ to the $D^{*0} \bar D^0$ threshold  and  the large decay rate to  $J/\psi \, \pi \pi \, (J/\psi \, \rho )$  compared  to  $J/\psi \, \pi \pi \pi \, (J/\psi \, \omega)$  \cite{Workman:2022ynf}, which
can be explained invoking a multiquark structure of $X(3872)$,  compact or molecular  (see the  discussions  in Refs.\cite{Brambilla:2019esw,Maiani:2022psl} and in references therein). On the other hand, the  large 
  $\Gamma(X(3872)\to \gamma \, \psi(2S))/\Gamma(X(3872) \to \gamma \, J/\psi)$ ratio \cite{ LHCb:2014jvf}  and data on the production  in  $\gamma \gamma^*$  interactions \cite{Belle:2020ndp} support the identification of   
$\chi_{c1}(3872)$ with the state $\chi_{c1}(2P)$ sitting on the $D^{*0} \bar D^0$ threshold \cite{Achasov:epj,Achasov:2022puf}. Experimental tests have been proposed  to help decipher the structure of the meson \cite{Colangelo:2007ph},  in  the vast literature on the subject.

Semileptonic $B_c$ decays to $\chi_{c1}(3872)$ can  provide us with  information on the nature of this state.
 The way is inspired by our analysis  in \cite{Colangelo:2022lpy} and is based on the systematic comparison of the $B_c$ transitions to the  $2P$ charmonia.   In a selected kinematical range the  $B_c$ and the P-wave charmonium matrix elements can be expressed as an expansion in the heavy quark relative three-velocity in the heavy hadrons, together with an expansion in the inverse heavy quark mass. In this range  the   $B_c$  to  the P-wave charmonia form factors (for  lowest lying  or radial excitations)  are related.   The  violation or confirmation  of such relations in $B_c \to \chi_{c1}(3872) \ell \bar \nu$ with respect to the  decays  to the other $2P$ charmonia  would  support the interpretation of $X(3872)$ as an exotic or a conventional state.\footnote{ The comparison  between  $B_c$ semileptonic and nonleptonic $B_c\to X(3872) \rho (a_1)$ decays, fixing the  $X(3872)$ polarization, has been proposed as a way to investigate the structure of this meson  \cite{Wang:2015rcz}. Semileptonic $B_c$ decays to $X(3872)$ have  been considered  in \cite{Hsiao:2016pml}.}

In this paper we focus on  the $B_c$ to the $P$-wave charmonia form factors, and compute their expressions in terms of universal functions at  NLO.
We classify the universal functions and  work out a set of  relations useful both to test calculations based on nonperturbative methods and in phenomenological analyses.
 In Sec.~\ref{sec1}  we parametrize  the  $B_c \to \chi_{c0,1,2}$  and $B_c \to h_c$ matrix elements of the quark currents appearing in a  generalized low-energy Hamiltonian governing the $b \to c \ell \bar \nu_\ell$ transition,  providing the definition of the set of form factors considered in our study.  For the sake of completeness,   in Appendix \ref{app:ff} we also report another parameterization  often employed in the literature, with  straightforward relations between the two. 
 We give the decay distributions of the four $B_c \to \chi_{c 0,1,2} (h_c) \ell \bar \nu_\ell$  modes, which can also be used  for $B_c$ decays to the  $2P$ charmonium resonances.
 In Sec.~\ref{sec2} we briefly describe the heavy quark expansion in QCD, we introduce the $(B_c, B_c^*)$ spin doublet and the   $(\chi_{c0,1,2}, h_c)$ spin 4-plet and the trace formalism  to express the relevant hadronic matrix elements. In Sec.~\ref{sec3} we write the form factors in terms of universal functions, with the  resulting formulae collected in Appendix \ref{app:universalff}.
 In Sec.~\ref{sec4} we present  relations among several form factors for single modes and for pairs of modes,  other relations being collected in Appendix \ref{app:relations}.
Applications to phenomenology are discussed in Sec.~\ref{sec:discussion}, for the  lowest lying $1P$ charmonia and for the $2P$ excitations in connection with 
the $\chi_{c1}(3872)$ issue. Then we conclude.
 
\section{$B_c \to \chi_{c 0,1,2}$  and $B_c \to h_c$ form factors}\label{sec1}
The semileptonic $B_c$ decays to the charmonium states, including the positive parity $\chi_{c0,1,2}$ and $h_c$, are governed by a low energy
 Hamiltonian with  general form
\bea
H_{\rm eff}^{b \to c \ell \bar \nu}&=& \frac{G_F}{\sqrt{2}} V_{cb} \Big[(1+\epsilon_V^\ell) \left({\bar c} \gamma_\mu  (1-\gamma_5) b \right)\left(  \bar \ell  \gamma^\mu  (1-\gamma_5)   \nu_{\ell } \right)
+  \epsilon_R^\ell \left({\bar c} \gamma_\mu (1+\gamma_5) b \right)\left( \bar \ell  \gamma^\mu  (1-\gamma_5) \nu_{\ell } \right)  \nn \\
&+& \epsilon_S^\ell \, ({\bar c} b) \left( {\bar \ell} (1-\gamma_5) \nu_{\ell } \right)
+ \epsilon_P^\ell \, \left({\bar c} \gamma_5 b\right)  \left({\bar \ell} (1-\gamma_5)\nu_{\ell } \right)  \label{hamil}  \\
&+& \epsilon_T^\ell \, \left({\bar c}  \sigma_{\mu \nu} (1-\gamma_5) b\right) \,\left( {\bar \ell}   \sigma^{\mu \nu} (1-\gamma_5) \nu_{\ell }\right)     \Big]   \nn
\eea
considering the full set of  $D=6$ semileptonic  $b \to c$ operators with left-handed neutrinos.
The general Hamiltonian \eqref{hamil} comprises   the Fermi constant $G_F$, the  element $V_{cb}$ of the CKM matrix and   the Standard Model operator
  ${\cal O}_{SM}=4 (\bar c_L \gamma^\mu b_L) \left( {\bar \ell_L} \gamma_\mu \nu_{\ell L}\right)$. It also
includes 
 the operators ${\cal O}_{R}=4 (\bar c_R \gamma^\mu b_R) \left( {\bar {\ell_L}} \gamma_\mu \nu_{\ell L}\right)$,  
  ${\cal O}_S=\left({\bar c} b\right)\left( {\bar \ell} (1-\gamma_5) \nu_\ell \right)$, 
 ${\cal O}_P=\left({\bar c} \gamma_5 b \right)\left( {\bar \ell} (1-\gamma_5)  \nu_\ell \right)$ and 
 ${\cal O}_T=\left({\bar c}  \sigma_{\mu \nu} (1-\gamma_5) b \right)\left( {\bar \ell} \sigma^{\mu \nu}  (1-\gamma_5)  \nu_\ell \right)$ arising in  extensions of SM, 
with  Wilson  coefficients $\epsilon^\ell_{V,R,S,P,T}$ in general  complex and  lepton-flavour dependent. The SM  Hamiltonian corresponds to  $\epsilon^\ell_i=0$.
 Eq.~\eqref{hamil} has been considered in connection with the $B \to D^{(*)} \tau \nu_\tau$, $B \to D^{(*)} \ell \nu_\ell$ LFU anomaly \cite{Becirevic:2016hea,Alonso:2016oyd,Colangelo:2016ymy,Jung:2018lfu,Colangelo:2018cnj,Murgui:2019czp,Alguero:2020ukk}. 

It is common to parametrize the $B_c \to \chi_{ci}$ ($i=0,1,2$) and $B_c \to h_c$  matrix elements  of the quark currents in  Eq.~\eqref{hamil}
 in terms of form factors as written in  Appendix \ref{app:ff} 
 \cite{Wang:2009mi,Zhu:2017lwi,Rui:2018kqr,Wang:2018duy}.  
  $p$ and $p^\prime$ are the four momenta of the initial and final meson,    $\epsilon$ ($\eta$)  the polarization vector (tensor) of the spin $1 $ (spin $2$) charmonium,   $q=p-p^\prime$ the   lepton pair momentum. 
We use $\epsilon^{0123}=-1$, hence $\sigma^{\mu \nu} \gamma_5 = - \frac{i}{2} \varepsilon^{\mu \nu \alpha \beta} \sigma_{\alpha \beta}$. 
In our analysis it is more convenient to use a parametrization of the matrix elements in terms  of the four-velocities of $B_c$ and of the 
charmonium state $C=\chi_{c0,1,2},h_c$:  $v=p/m_{B_c}$ and $v^\prime=p^\prime/m_{C}$,  with $w=v \cdot v^\prime$. They  are defined as follows.\\

\noindent $B_c \to \chi_{c0}$:
\bea
\langle \chi_{c0}(v^\prime) | \bar{c}  \gamma_\mu  \gamma_5 b | B_c(v) \rangle &=& \sqrt{m_{\chi_{c0}} \, m_{B_c}} \,\big[g_+(w)(v+v^\prime)_\mu+g_-(w)(v-v^\prime)_\mu\big]
 \nn  \\
\langle \chi_{c0}(v^\prime) | \bar{c}  \gamma_5  b | B_c(v) \rangle &=& \sqrt{m_{\chi_{c0}} \, m_{B_c}} \,\,g_P(w) \label{ff:chic0}\\
\langle \chi_{c0}(v^\prime) | \bar{c}  \sigma_{\mu\nu}  b | B_c(v) \rangle &=&\sqrt{m_{\chi_{c0}} \, m_{B_c}} \,\, g_T(w)\,
\epsilon_{\mu \nu \alpha \beta}v^\alpha v^{\prime \beta}
\nn
\eea
\noindent $B_c \to \chi_{c1}$:
\bea
\langle \chi_{c1}(v^\prime,\epsilon) | \bar{c}  \gamma_\mu  b | B_c(v) \rangle &=&i \,\sqrt{m_{\chi_{c1}} \, m_{B_c}} \,\Big[g_{V_1}(w)\epsilon_{ \mu}^*\nn \\&+&(\epsilon^* \cdot v)[g_{V_2}(w) (v+v^\prime)_\mu+g_{V_3}(w) (v-v^\prime)_\mu ] \Big]
\nn \\
\langle \chi_{c1}(v^\prime,\epsilon) | \bar{c}  \gamma_\mu  \gamma_5 b | B_c(v) \rangle &=& \sqrt{m_{\chi_{c1}} \, m_{B_c}} \, g_A(w)\,\epsilon_{\mu  \alpha \beta \sigma }\epsilon^{* \alpha}v^\beta v^{\prime \sigma}
 \nn  \\
\langle \chi_{c1}(v^\prime,\epsilon) | \bar{c}    b | B_c(v) \rangle &=& i\,\sqrt{m_{\chi_{c1}} \, m_{B_c}} \,g_S(w) \,(\epsilon^* \cdot v) \label{ff:chic1}  \\
\langle \chi_{c1}(v^\prime,\epsilon) | \bar{c}  \sigma_{\mu\nu}  b | B_c(v) \rangle &=&\sqrt{m_{\chi_{c1}} \, m_{B_c}} \Big[g_{T_1}(w)\big(\epsilon_{\mu}^*(v+v^\prime)_\nu-\epsilon_{\nu}^*(v+v^\prime)_\mu\big) \nn \\
&+&g_{T_2}(w)\big(\epsilon_{\mu}^*(v-v^\prime)_\nu-\epsilon_{\nu}^*(v-v^\prime)_\mu\big)
\nn \\
&+&g_{T_3}(w)(\epsilon^* \cdot v)(v_\mu v^\prime_\nu-v_\nu v_\mu^\prime)\Big]
\nn
\eea
\noindent $B_c \to \chi_{c2}$:
\bea
\langle \chi_{c2}(v^\prime,\eta) | \bar{c}  \gamma_\mu  b | B_c(v) \rangle &=&\sqrt{m_{\chi_{c2}} \, m_{B_c}} \,i\,k_V(w)\,
\epsilon_{\mu \alpha \beta \sigma} \eta^{*\alpha \tau}v_\tau v^\beta v^{\prime \sigma} \nn \\
\langle \chi_{c2}(v^\prime,\eta) | \bar{c}  \gamma_\mu  \gamma_5 b | B_c(v) \rangle &=&\sqrt{m_{\chi_{c2}} \, m_{B_c}} \,
\big[k_{A_1}(w) \,\eta_{\mu \alpha}^*v^\alpha+\eta_{ \alpha \beta}^*v^\alpha v^\beta \left(k_{A_2}(w) v_\mu +k_{A_3}(w)v^\prime_\mu \right) \big]
\nn \\
\langle \chi_{c2}(v^\prime,\eta) | \bar{c}   \gamma_5 b | B_c(v) \rangle &=&\sqrt{m_{\chi_{c2}} \, m_{B_c}} k_P(w)\,\eta_{ \alpha \beta}^*v^\alpha v^\beta   \label{ff:chic2}\\
\langle \chi_{c2}(v^\prime,\eta) | \bar{c}   \sigma_{\mu\nu} \gamma_5  b | B_c(v) \rangle &=&i \sqrt{m_{\chi_{c2}} \, m_{B_c}} \, \big[ k_{T_1}(w) ( {\eta^*}_\mu^\alpha v_\alpha v_\nu - {\eta^*}_\nu^\alpha v_\alpha v_\mu) + \nn \\
&+& k_{T_2}(w) ( {\eta^*}_\mu^\alpha v_\alpha v'_\nu - {\eta^*}_\nu^\alpha v_\alpha v'_\mu) + k_{T_3}(w) \eta_{ \alpha \beta}^*v^\alpha v^\beta ( v_\mu v'_\nu - v_\nu v'_\mu ) \big]
\nn
\eea
\noindent $B_c \to h_c$:
\bea
\langle h_c(v^\prime,\epsilon) | \bar{c}  \gamma_\mu  b | B_c(v) \rangle &=&\sqrt{m_{h_c} \, m_{B_c}} \,\Big[f_{V_1}(w)\epsilon_{\mu}^*\nn \\ &+&(\epsilon^* \cdot v)\big(f_{V_2}(w)(v+v^\prime)_\mu+f_{V_3}(w)(v- v^\prime)_\mu \big) \Big]
\nn \\
\langle h_c(v^\prime,\epsilon) | \bar{c}  \gamma_\mu  \gamma_5 b | B_c(v) \rangle &=& i \sqrt{m_{h_{c}} \, m_{B_c}} \, f_A(w)\,\epsilon_{\mu  \alpha \beta \sigma }\epsilon^{* \alpha}v^\beta v^{\prime \sigma}
\nn \\
\langle h_c(v^\prime,\epsilon) | \bar{c}   b | B_c(v) \rangle &=&\sqrt{m_{h_c} \, m_{B_c}} \,(\epsilon^* \cdot v)f_S(w) \label{ff:hc}
 \\
\langle h_c(v^\prime,\epsilon) | \bar{c}  \sigma_{\mu\nu} b | B_c(v) \rangle &=&i \,\sqrt{m_{h_c} \, m_{B_c}} \Big[f_{T_1}(w)\big(\epsilon_{\mu}^*(v+v^\prime)_\nu-\epsilon_{\nu}^*(v+v^\prime)_\mu\big)\nn \\
&+&f_{T_2}(w)\big(\epsilon_{\mu}^*(v-v^\prime)_\nu-\epsilon_{\nu}^*(v-v^\prime)_\mu\big)
\nn \\
&+& \,f_{T3}(w)\,(\epsilon^* \cdot v)\left(v_\mu v^\prime_\nu-v_\nu v^\prime_\mu \right) \Big] .
\nn \eea
The  decay distributions  governed by the Hamiltonian \eqref{hamil} can be written in the form:
\bea
\frac{d\Gamma (B_c \to C \ell \bar \nu)}{dw}&=&{\tilde \Gamma} \Big\{ |1+\epsilon_V|^2 \,\frac{d\Gamma}{dw}^{SM}+|\epsilon_R|^2\frac{d\Gamma}{dw}^{R}+|\epsilon_X|^2\frac{d\Gamma}{dw}^{X}+|\epsilon_T|^2\frac{d\Gamma}{dw}^{T} \nn \\&+&
2 \, {\rm Re}\left[\epsilon_R(1+\epsilon_V^* )\right] \frac{d\Gamma}{dw}^{SMR} 
+2 \, {\rm Re}\left[\epsilon_X(1+\epsilon_V^* )\right] \frac{d\Gamma}{dw}^{SMX}  \nn \\
&+&2 \, {\rm Re}\left[\epsilon_T(1+\epsilon_V^* )\right]\frac{d\Gamma}{dw}^{SMT} 
+ 2 \, {\rm Re}\left[\epsilon_R \epsilon_T^* \right] \frac{d\Gamma}{dw}^{RT} \label{dGdw}  \\ 
&+&2 \, {\rm Re}\left[\epsilon_X \epsilon_R^* \right] \frac{d\Gamma}{dw}^{XR}+
2 \, {\rm Re}\left[\epsilon_X \epsilon_T^* \right] \frac{d\Gamma}{dw}^{XT}\Big\}, \,\,\,\, \nn 
\eea
with $X=P$ in case of  $\chi_{c0},\,\chi_{c2}$, and   $X=S$ in  case of $\chi_{c1},\,h_c$. 
In Eq.~\eqref{dGdw}  we define
\be 
{\tilde \Gamma}=\displaystyle\frac{G_F^2 |V_{cb}|^2m_{B_c}^5\,r^3}{48 \pi^3}\sqrt{w^2-1}\left(1-\frac{{\hat m}_\ell^2}{1 + r^2 - 2 r w} \right)^2 \label{eq:gammatilde}
\ee
with $r=m_C/m_{B_c}$ ($C=\chi_{c0,c1,c2},h_c$) and ${\hat m}_\ell=m_\ell/m_{B_c}$.
For the four decay modes  the expressions of the  functions in \eqref{dGdw} in terms of the form factors in \eqref{ff:chic0}-\eqref{ff:hc} are as follows.

\begin{itemize}
\item
$C=\chi_{c0}$:
\bea
&&\frac{d\Gamma}{dw}^{SM}=\frac{d\Gamma}{dw}^{R}=-\frac{d\Gamma}{dw}^{SMR}  \nn \\
&=&[g_+(w)]^2 \, (w+1) \Big[(w-1)(1+r)^2+\frac{{\hat m}_\ell^2}{1+r^2-2rw}\Big( (2w+1)(1+r^2)-2r(w+2) \Big) \Big]  \nn \\
&+&[g_-(w)]^2 \, (w-1) \Big[(w+1)(1-r)^2+\frac{{\hat m}_\ell^2}{1+r^2-2rw} \Big( (2w-1)(1+r^2)+2r(w-2) \Big) \Big] \quad \nn \\
&-&2g_+(w)g_-(w)\,(w^2-1)(1-r^2)\left(1+\frac{2{\hat m}_\ell^2}{1+r^2-2rw} \right) 
\eea
\bea
\frac{d\Gamma}{dw}^{P}&=& \frac{3}{2} [g_P(w)]^2 (1+r^2-2rw)  \\
\frac{d\Gamma}{dw}^{T}&=& 8 [g_T(w)]^2 \left(1+r^2-2rw+2{\hat m}_\ell^2 \right)  (w^2-1) 
\eea
\bea
\frac{d\Gamma}{dw}^{SMP}&=&-\frac{d\Gamma}{dw}^{PR} 
= \frac{3}{2} g_P(w) {\hat m}_\ell\,\Big[(w-1)(1+r)g_-(w)-(w+1)(1-r)g_+(w)\Big]  \\
\frac{d\Gamma}{dw}^{SMT}&=&-\frac{d\Gamma}{dw}^{RT} 
=-6 g_T(w)\,{\hat m}_\ell(w^2-1)\,\Big[(1+r)g_+(w)-(1-r)g_-(w)\Big] \,\, .
\eea 
\item
$C=\chi_{c1}$:
\bea
&&
\frac{d\Gamma}{dw}^{SM}=\frac{d\Gamma}{dw}^{R}\nn \\
&&=[g_{V_1}(w)]^2\,\Big[3(w-r)^2-2(w^2-1)+\frac{{\hat m}_\ell^2}{2(1+r^2-2rw)}\Big(3(w-r)^2+w^2-1 \Big) \Big]
\nn \\
&&+[g_{V_2}(w)]^2\,(w^2-1)(w+1)\Big[(1+r)^2(w-1)+\frac{{\hat m}_\ell^2}{1+r^2-2rw}\Big((2w+1)(1+r^2)-2r(w+2) \Big) \Big]\nn \\
&&+[g_{V_3}(w)]^2\,(w^2-1)(w-1)\Big[(1-r)^2(w+1)+\frac{{\hat m}_\ell^2}{1+r^2-2rw}\Big((2w-1)(1+r^2)+2r(w-2) \Big) \Big]\nn \\
&&+[g_A(w)]^2\,(w^2-1)\Big(2(1+r^2-2rw)+{\hat m}_\ell^2 \Big)  \\
&&+g_{V_1}(w)\,g_{V_2}(w)\,(w^2-1)\Big[2(1+r)(w-r)-\frac{{\hat m}_\ell^2}{1+r^2-2rw}\Big(-3+r^2-4w+2r(w+2) \Big) \Big]\nn \\
&&-g_{V_1}(w)\,g_{V_3}(w)\,(w^2-1)\Big[2(1-r)(w-r)+\frac{{\hat m}_\ell^2}{1+r^2-2rw}\Big(-3+r^2+4w+2r(w-2) \Big) \Big]\nn \\
&&-2g_{V_2}(w)\,g_{V_3}(w)\,(w^2-1)^2(1-r^2)\left(1+\frac{2 {\hat m}_\ell^2}{1+r^2-2rw} \right) \nn 
\eea
\bea
\frac{d\Gamma}{dw}^{SMR}&=&
\frac{d\Gamma}{dw}^{SM}-2[g_A(w)]^2\,(w^2-1)\Big(2(1+r^2-2rw)+{\hat m}_\ell^2 \Big) \\
\frac{d\Gamma}{dw}^{S}&=& \frac{3}{2} [g_S(w)]^2 (1+r^2-2rw)(w^2-1) 
\\
\frac{d\Gamma}{dw}^{T}&=&8\left(1+\frac{2 {\hat m}_\ell^2}{1+r^2-2rw} \right)\, \nn \\
&\times& \Big\{[g_{T_1}(w)]^2\,(w+1)\Big[(5w+1)(1+r^2)-2r(w^2+w+4) \Big]
\nn \\
&+&\,[g_{T_2}(w)]^2\,(w-1)\Big[(5w-1)(1+r^2)-2r(w^2-w+4) \Big]
\nn \\
&+&\,[g_{T_3}(w)]^2\,(w^2-1)^2(1+r^2-2rw)  \\
&+&2\, g_{T_1}(w)\,g_{T_2}(w)\,(w^2-1)(5r^2-2rw-3) \nn \\
&-&2\, g_{T_1}(w)\,g_{T_3}(w)\,(w^2-1)(w+1)(1+r^2-2rw) \nn \\
&-&2\, g_{T_2}(w)\,g_{T_3}(w)\,(w^2-1)(w-1)(1+r^2-2rw) \Big\} \nn 
\eea
\bea
\frac{d\Gamma}{dw}^{SMS}=\frac{d\Gamma}{dw}^{SR} &=& \frac{3}{2}
g_S(w) {\hat m}_\ell(w^2-1)  \nn \\
&\times&\Big[g_{V_1}(w)+(w+1)(1-r)g_{V_2}(w)-(w-1)(1+r)g_{V_3}(w)\Big] 
\eea
\bea
&&
\frac{d\Gamma}{dw}^{SMT}=\nn \\ 
&& 6\,{\hat m}_\ell \Big\{-g_{T_1}(w)(w+1)\Big[(2-3r+w)\,g_{V_1}(w)+(w^2-1)\big[(1+r)g_{V_2}(w)-(1-r)g_{V_3}(w) \big] \Big]\nn \\
&&
-g_{T_2}(w)(w-1)\Big[(-2-3r+w)\,g_{V_1}(w)+(w^2-1)\big[(1+r)g_{V_2}(w)-(1-r)g_{V_3}(w) \big] \Big]\nn \\
&&
+g_{T_3}(w)(w^2-1)\Big[(w-r)\,g_{V_1}(w)+(w^2-1)\big[(1+r)g_{V_2}(w)-(1-r)g_{V_3}(w) \big] \Big] \\
&&+2(w^2-1)g_A(w)\Big[(1+r)g_{T_1}(w)-(1-r)g_{T_2}(w)\Big]\Big\} \nn
\eea
\be
\frac{d\Gamma}{dw}^{RT}=
\frac{d\Gamma}{dw}^{SMT}-24\,{\hat m}_\ell (w^2-1)g_A(w)\Big[(1+r)g_{T_1}(w)-(1-r)g_{T_2}(w)\Big]  \,\, .
\ee
\item
$C=\chi_{c2}$:
\bea
\frac{d\Gamma}{dw}^{SM}&=&\frac{d\Gamma}{dw}^{R}
=\frac{w^2-1}{6} \Big\{[k_V(w)]^2\,3\left(2+\frac{{\hat m}_\ell^2}{1+r^2-2rw} \right)(w^2-1)(1+r^2-2rw)\nn \\
&+&[k_{A_1}(w)]^2 \Big[2(3+5r^2-10rw+2w^2)+\frac{{\hat m}_\ell^2}{1+r^2-2rw}\,(-3+5r^2-10rw+8w^2) \Big] \nn \\
&+&[k_{A_2}(w)]^2 2(w^2-1)\Big[2r^2(w^2-1)+\frac{{\hat m}_\ell^2}{1+r^2-2rw}\,\Big(3-6rw+r^2(4w^2-1) \Big)\Big] \nn \\
&+&[k_{A_3}(w)]^2 2(w^2-1)\Big[2(w^2-1)+\frac{{\hat m}_\ell^2}{1+r^2-2rw}\,\Big(-1+3r^2-6rw+4w^2) \Big)\Big] \nn \\
&+&4(w^2-1) \Big[k_{A_1}(w) \,k_{A_2}(w) \, \big[2r(w-r)+\frac{{\hat m}_\ell^2}{1+r^2-2rw}\,(3-r^2-2rw) \big]  \\
& -&k_{A_2}(w) \,k_{A_3}(w) \,\big[-2r(w^2-1)+\frac{{\hat m}_\ell^2}{1+r^2-2rw}\, \big[2r(w^2+2)-3w(1+r^2) \big] \nn \\
&+&2k_{A_1}(w) \,k_{A_3}(w) \,\left(1+\frac{2{\hat m}_\ell^2}{1+r^2-2rw} \right)(w-r)  \Big] \Big\} \nn
\\
\frac{d\Gamma}{dw}^{SMR}&=&-\frac{d\Gamma}{dw}^{SM}+[k_V(w)]^2 \,\Big(2(1+r^2-2rw)+{\hat m}_\ell^2 \Big)(w^2-1)^2
\eea
\bea
\frac{d\Gamma}{dw}^{P}&=&[k_P(w)]^2\,(w^2-1)^2(1+r^2-2rw)
\\
\frac{d\Gamma}{dw}^{T}&=&\frac{8}{3}\left(1+\frac{2{\hat m}_\ell^2}{1+r^2-2rw} \right)(w^2-1)\Big\{[k_{T_1}(w)]^2 \Big[(3+2w^2)(1-2rw)+r^2(8w^2-3)\Big] \nn \\
&+&[k_{T_2}(w)]^2 \,(-1+5r^2-10rw+6w^2)+[k_{T_3}(w)]^2 \,2(w^2-1)^2(1+r^2-2rw) \quad  \nn \\
&-&2k_{T_1}(w)\,k_{T_2}(w)\,\big[6r-5w(1+r^2)+4rw^2 \big] \\ 
&-&4(w^2-1)(1+r^2-2rw)k_{T_3}(w) \big[w\,k_{T_1}(w)+k_{T_2}(w)\big] \Big\} \nn
\eea
\bea
\frac{d\Gamma}{dw}^{SMP}&=&-\frac{d\Gamma}{dw}^{PR} \nn \\
&=&-k_P(w)\,{\hat m}_\ell (w^2-1)^2 \Big[k_{A_1}(w)+(1-rw)k_{A_2}(w)+(w-r)k_{A_3}(w) \Big] 
\eea
\bea
&&
\frac{d\Gamma}{dw}^{SMT}=2{\hat m}_\ell (w^2-1) \nn  \\
&&\times \Big\{k_{T_1}(w) \Big[(3 - 5 r w + 2 w^2)\, k_{A_1}(w) + ( w^2-1) \Big(2 r w \,k_{A_2}(w) + 2 w \,k_{A_3}(w) + 3 r k_V(w)\Big) \Big] \nn \\
   &&+k_{T_2}(w) \Big[5 (w-r) \,k_{A_1}(w) + ( w^2-1) \Big(2 r \,k_{A_2}(w) + 2\, k_{A_3}(w) + 3 k_V(w)\Big)\Big]  \\
   &&-2k_{T_3}(w)\,(w^2-1) \Big[(w-r) k_{A_1}(w) + ( w^2-1) \Big(r \,k_{A_2}(w) + k_{A_3}(w)\Big) \Big] \Big \} \nn
   \\
   &&\frac{d\Gamma}{dw}^{RT}=-\frac{d\Gamma}{dw}^{SMT}+12{\hat m}_\ell (w^2-1)^2k_V(w) \Big[r\, k_{T_1}(w) + k_{T_2}(w)\Big] \,\, .
   \eea
\item
$C=h_c$:

Looking at the matrix elements for $\chi_{c1}$ and $h_c$ in \eqref{ff:chic1} and \eqref{ff:hc}, 
the distributions for $h_c$ are obtained from those of $\chi_{c1}$ replacing
\bea
&&g_{V_i}g_{V_j} \to f_{V_i} f_{V_j} , \quad  g_A^2 \to  f_A^2,  \quad g_S^2 \to  f_S^2 , \quad g_{T_i}g_{T_j} \to f_{T_i} f_{T_j} , \nn \\
&&g_{S}g_{V_i} \to f_{S} f_{V_i}  , \quad g_{V_i}g_{T_j} \to - f_{V_i} f_{T_j} , \quad g_{A}g_{T_i} \to f_{A} f_{T_i} ,
\eea
with $i,j=1,2,3$.
\end{itemize}
The above expressions also hold for $B_c$ decays to the $2P$ (and higher)  charmonium resonances. They are the subject of our analysis  together with the form factors in Eqs.~(\ref{ff:chic0})-(\ref{ff:hc}).

\section{Form factors in the effective theory}\label{sec2}
As discussed  in \cite{Colangelo:2022lpy}, we aim at expressing the form factors   in the effective theory of QCD resulting by an expansion in the inverse heavy quark mass. 
The  heavy quark QCD field $Q(x)$ with mass $m_Q$  is written, with the notations in \cite{Aebischer:2021ilm}, as
\be
Q(x)=  e^{-i \, m_Q v \cdot x} \psi(x)=e^{-i \, m_Q v \cdot x} \Big( \psi_+(x)+\psi_-(x) \Big) \label{eq:Q}
\ee
with  $\psi_\pm= P_\pm \psi(x)$ and  $\dd P_\pm=\frac{1\pm\slashed{v}}{2}$. $\psi_+$  is the positive energy component of the field,  $v$ is  the heavy meson (quarkonium) 4-velocity with $v^2=1$. Hence, $Q(x)$ is expressed as
\be
Q(x)= e^{-i \, m_Q v \cdot x} \left( 1+\frac{i \slashed{D}_\perp}{2 m_Q} +\frac{(-i v \cdot D)}{2 m_Q}  \frac{i \slashed{D}_\perp}{2 m_Q} + \dots    \right)\psi_+(x) , \label{eq:10}
\ee
with $D_{\perp \mu}= D_\mu -(v \cdot D) v_\mu$. 
In the rest frame $v=(1,\vec 0)$ we have  $v \cdot D=D_t$ and $D_{\perp \mu}=(0, D_i)$.
The  QCD Lagrangian is expressed in terms of $\psi_+$ as an expansion,
\bea
{\cal L}_{QCD}&=&{\bar \psi}_+(x) \left(i v \cdot D+\frac{(iD_\perp)^2}{2m_Q} +\frac{g}{4 m_Q}\sigma \cdot G_\perp  +\frac{i \slashed{D}_\perp}{2 m_Q} \frac{(-i v \cdot D)}{2 m_Q}  ( i \slashed{D}_\perp)+ \dots \right) \psi_+(x) \nn \\
&=& {\cal L}_0+{\cal L}_1 + \dots \,\,\, , \label{eq:lagexp}
\eea
with $G_{\perp \mu \nu}=(g_{\mu \alpha}-v_\mu v_\alpha)(g_{\nu \beta}-v_\nu v_\beta)G^{\alpha \beta}$. 

Nonrelativistic QCD provides us with power counting of the  operators in terms of    the relative heavy quark 3-velocity in the hadron rest frame   $ \tilde v = | \vec{\tilde v}|\ll 1$, starting from $D_t \sim {\tilde v}^2$, $D_\perp \sim {\tilde v}$ and  $\psi_+\sim  \tilde v^{3/2}$ \cite{Bodwin:1994jh}. Hence,
the second  term in Eq.~\eqref{eq:10} is  ${\cal O}(\tilde v \times \tilde v^{3/2})$,  the third one is  ${\cal O}(\tilde v^3\times \tilde v^{3/2})$, and so on.  In the following we  omit  the power $\tilde v^{3/2}$ for each quark field  in the power counting of the various operators. Then,
the  chromoelectric  field components $E_i=G_{0i}$ and  the chromomagnetic ones $B_i=\displaystyle\frac{1}{2}\epsilon_{ijk}G^{jk}$ are  ${\cal O} (\tilde v^3)$ and ${\cal O}( \tilde v^4)$, respectively \cite{Lepage:1992tx}. 

The first two terms  in Eq.~\eqref{eq:lagexp} are   ${\cal O}(\tilde v^2)$. They give  the leading order effective Lagrangian 
\bea
{\cal L}_0&=& {\bar \psi}_+(x) \Big(i v \cdot D+\frac{(iD_\perp)^2}{2m_Q}\Big) \psi_+(x) .
\eea
The third and fourth term in Eq.~\eqref{eq:lagexp} are  ${\cal O}(\tilde v^4)$. They correspond to 
the NLO term
\bea
{\cal L}_1&=&{\cal L}_{1,1}+{\cal L}_{1,2} \,\, , \label{eq:l1}
\eea
where
\bea
 {\cal L}_{1,1}&=&{\bar \psi}_+(x) \frac{g \sigma \cdot G_\perp}{4 m_Q} \psi_+(x) \nn \\
 {\cal L}_{1,2}&=&\bar \psi_+(x) \frac{i \slashed{D}_\perp}{2 m_Q} \frac{(-i v \cdot D)}{2 m_Q}  ( i \slashed{D}_\perp) \psi_+(x) .
 \eea
${\cal L}_{1,2}$ can be written as
 \bea
  {\cal L}_{1,2}&=&-\frac{1}{4m_Q^2}\Bigg({\bar \psi}_+(x) \left( -\frac{(i D_\perp)^4}{2m_Q}\right) \psi_+(x) +{\bar \psi}_+(x)\frac{g}{2}\sigma \cdot G_\perp \left( -\frac{(i D_\perp)^2}{2m_Q}\right) \psi_+(x) \nn \\
  &+&ig v^\alpha {\bar \psi}_+(x) iD_\perp^\sigma G_{ \alpha \sigma} \psi_+(x) +g v^\alpha {\bar \psi}_+(x)i D_{\perp \tau} \sigma^{\tau \sigma}G_{ \alpha \sigma}\psi_+(x) \Bigg) \\
  &=&{\cal L}_{1,2}^{(1)}+{\cal L}_{1,2}^{(2)}+{\cal L}_{1,2}^{(3)}+{\cal L}_{1,2}^{(4)} , \nn
  \eea
with ${\cal L}_{1,2}^{(2)}$  of higher order in $\tilde v$. 
${\cal L}_1$  can be arranged  as
 \be
 {\cal L}_1={\cal L}_1^A+{\cal L}_1^B  \,\,  \label{L1}\\
 \ee
 with
 \bea
 {\cal L}_1^A&=&{\cal L}_{1,1}+{\cal L}_{1,2}^{(4)} =\frac{1}{4m_Q}{\bar \psi}_+(x)A_{\tau \sigma}\sigma^{\tau \sigma} \psi_+(x) \label{nonlocal1}\\
 {\cal L}_1^B &=&{\cal L}_{1,2}^{(1)}+{\cal L}_{1,2}^{(3)}=\frac{1}{4m_Q^2}{\bar \psi}_+(x)B \psi_+(x) ,  \label{nonlocal2}
 \eea
 factorizing in \eqref{nonlocal1} the leading $1/m_Q$ power.
 
An analogous expansion describes  the antiquark: $Q(x)$  is written as
\be
Q(x)=  e^{i \, m_Q v \cdot x} X(x)=e^{i \, m_Q v \cdot x} \Big( X_+(x)+X_-(x) \Big)
\ee
giving 
\be
{\cal L}_{QCD}={\overline X}_-(x) \left(- i v \cdot D +\frac{(iD_\perp)^2}{2m_Q}  + \dots \right) X_-(x) \label{lag-qbar}.
\ee

To express the heavy meson form factors in the effective QCD theory, we expand
the weak current involving two heavy quarks $\bar Q^\prime \Gamma Q$, with $\Gamma$  a generic Dirac matrix:
\bea
{\bar Q^\prime(x)}\Gamma Q(x)&=&{\bar \psi}^\prime_+(x) \Big( 1-\frac{i {\overleftarrow {\slashed D}'}_\perp }{2 m_{Q^\prime}} -\frac{1}{4m_{Q^\prime}^2}(i {\overleftarrow {\slashed D}'}_\perp )(i v^\prime \cdot {\lvecD}) +\dots   \Big)  \nn \\   &&\Gamma  \Big( 1+\frac{i {\overrightarrow {\slashed D}}_\perp}{2 m_Q} +\frac{1}{4m_Q^2}(-i v \cdot {\rvecD}) i {\overrightarrow {\slashed D}}_\perp  + \dots \Big) \psi_+(x)  \,\, ,
\eea
where $D_{\perp \mu}^\prime= D_\mu -(v^\prime \cdot D) v_\mu^\prime$. 
Up to  ${\cal O}(1/m_Q^2)$ the expansion reads:
\be
{\bar Q^\prime}(x)\Gamma Q(x)=J_0+\Big(\frac{J_{1,0}}{2 m_Q}+\frac{J_{0,1}}{2 m_{Q^\prime}}\Big)+\Big(-\frac{J_{2,0}}{4 m_Q^2}-\frac{J_{0,2}}{4 m_{Q^\prime}^2}+\frac{J_{1,1}}{4 m_Q m_{Q^\prime}} \Big) , \label{current}
 \ee
 with 
\bea
J_0 &=& {\bar \psi}_+' \Gamma \psi_+  \nn \\
J_{1,0} &=& {\bar \psi}_+' \Gamma i {\overrightarrow {\slashed D}}_\perp \psi_+ \nn \\
J_{0,1} &=& {\bar \psi}_+' \left( -i {\overleftarrow {\slashed D}'}_\perp \right) \Gamma \psi_+ \nn \\
J_{2,0} &=& {\bar \psi}_+' \Gamma \left(i v \cdot \rvecD \right) i {\overrightarrow {\slashed D}}_\perp\psi_+ \\
J_{0,2} &=& {\bar \psi}_+'  i {\overleftarrow {\slashed D}'}_\perp \left( i v' \cdot \lvecD \right) \Gamma \psi_+ \nn \\
J_{1,1} &=& {\bar \psi}_+' \left( -i {\overleftarrow {\slashed D}'}_\perp \right)\Gamma \left( i {\overrightarrow {\slashed D}}_\perp \right) \psi_+ . \nn
\eea
Eq.~\eqref{current}  comprises  terms up to ${\cal O}(\tilde v^3)$.  ${\cal O}(1/m_Q^3)$ terms with three derivatives have not been included, even though they can be of the same order in  $\tilde v$ of some terms in the 
corrections  discussed in the following:  we assume that their contribution is small.  

 The hadronic matrix elements of the operators in Eq.~\eqref{current} can be written using the trace formalism \cite{Falk:1991nq}. 
For states comprising a heavy quark and a heavy antiquark, spin symmetry is expected to hold in the preasymptotic mass range which includes beauty and charm, outside the QCD Coulombic regime \cite{Jenkins:1992nb,Casalbuoni:1992yd}.  Hence, the two states $(B_c, B_c^*)$ and  the four states $(\chi_{c0,1,2}, h_c)$ can be organized in a negative parity spin doublet  and in a  positive parity spin 4-plet.  They are described by $4 \times 4$ matrices, for   the spin doublet  \cite{Jenkins:1992nb} 
\be
\mathcal{M}(v)  = P_+(v) \, \left[ B_c^{*\mu}  \, \gamma_\mu - B_c \, \gamma_5 \right] \, P_-(v) \;, \label{doppietto}
\ee
for the spin 4-plet  \cite{Casalbuoni:1992fd}
\be
\mathcal{M}'^\mu(v')  = P_+(v') \, \bigg[ \chi_{c2}^{\mu\nu} \, \gamma_\nu + \frac{1}{\sqrt{2}} \, \chi_{c1,\gamma} \, \epsilon^{\mu\alpha\beta\gamma} \, v'_\alpha \, \gamma_\beta \,  + \frac{1}{\sqrt{3}} \, \chi_{c0} \, ( \gamma^\mu - v'^\mu ) + h_c^\mu \, \gamma_5 \bigg] \, P_-(v') \label{multipletto} 
\ee
with the condition 
\begin{equation}
v'_\mu \, \mathcal{M}'^\mu = 0 \;. \label{eq:cond}
\end{equation}
The normalization is  $\sqrt{m_M}$, with $M$  a  meson in the spin multiplet. Radial excitations belong to multiplets analogous to \eqref{doppietto} and \eqref{multipletto}. The  trace formalism has been used  to obtain   the effective Lagrangians governing  strong and radiative heavy quarkonium transitions  in the soft-exchange approximation \cite{Casalbuoni:1992yd,Casalbuoni:1992fd,DeFazio:2008xq}. We apply it to express the various form factors in terms of (universal) functions independent of the particular decay mode.

\section{Form factors in terms of universal functions}\label{sec3}

Applying the trace formalism, the matrix elements of the leading order term of the expansion of the quark  current \eqref{current} are written as
\begin{equation}
\label{deltaLO}
\braket{ M'(v') | J_0 | M(v) } = - \Xi(w) \, v_\mu \, \tr \big[ \overline{\mathcal{M}}'^\mu \, \Gamma \, \mathcal{M} \big] \;
\end{equation}
involving the single universal function $\Xi(w)$.

At $1 / m_Q$   the relevant matrix elements are parametrized as 
\begin{subequations}
\begin{align}
\label{dx_der}
\braket{ M'(v') | \bar{\psi}'_+ \, \Gamma \, i \, \Dright_\alpha \, \psi_+ | M(v) } = - \tr \big[ \Sigma_{\mu\alpha}^{(b)} \, \overline{\mathcal{M}}'^\mu \, \Gamma \, \mathcal{M} \big] \;, \\
\label{sx_der}
\braket{ M'(v') | \bar{\psi}'_+ \, ( - i \, \Dleft_\alpha ) \, \Gamma \, \psi_+ | M(v) } = - \tr \big[ \Sigma_{\mu\alpha}^{(c)} \, \overline{\mathcal{M}}'^\mu \, \Gamma \, \mathcal{M} \big] \;.
\end{align}
\end{subequations}
The general expression of the functions $\Sigma_{\mu\alpha}^{(Q)}$  is

\begin{equation}
\label{Sigmas}
\Sigma_{\mu\alpha}^{(Q)} = \Sigma_1^{(Q)} \, g_{\mu\alpha} + \Sigma_2^{(Q)} \, v_\mu \, v_\alpha + \Sigma_3^{(Q)} \, v_\mu \, v'_\alpha \, + \Sigma_4^{(Q)} \, v_\mu \, \gamma_\alpha + \Sigma_5^{(Q)} \, \gamma_\mu \, v_\alpha + \Sigma_6^{(Q)} \, \gamma_\mu \, v'_\alpha + \Sigma_7^{(Q)} \, i \, \sigma_{\mu\alpha} .
\end{equation}
The terms proportional to $v'^\mu$ vanish due to \eqref{eq:cond}.
$\Sigma_i^{(Q)}(w)$ $(i=1,\dots 7)$ are the universal functions at this order.
Upon integration by parts, relations can be worked out.  Using 
\begin{equation}
i \, \partial_\alpha \, ( \bar{\psi}'_+ \, \Gamma \, \psi_+ ) = \bar{\psi}'_+ \, i \, \Dleft_\alpha \, \Gamma \, \psi_+ + \bar{\psi}'_+ \, \Gamma \, i \, \Dright_\alpha \, \psi_+ \;,
\end{equation}
we obtain:
\begin{equation}
\label{partial}
- \big( \tilde{\Lambda} \, v_\alpha - \tilde{\Lambda}' \, v'_\alpha  \big) \, \Xi \, v_\mu \, \tr \big[ \overline{\mathcal{M}}'^\mu \, \Gamma \, \mathcal{M} \big] = \tr \big[ \Sigma^{(c)}_{\mu\alpha} \, \overline{\mathcal{M}}'^\mu \, \Gamma \, \mathcal{M} \big] - \tr \big[ \Sigma_{\mu\alpha}^{(b)} \, \overline{\mathcal{M}}'^\mu \, \Gamma \, \mathcal{M} \big] \; .
\end{equation}
 For the  $Q{\bar Q}^\prime$ mesons in the initial and final state  the parameters $\tilde \Lambda$,    $\tilde \Lambda^\prime$ are given by
\be 
\tilde \Lambda=m_H-m_Q-m_{{\bar Q}^\prime}  \label{eq:lambda}
\ee
and analogously for $\tilde \Lambda^\prime$.
The  relations follow:
\begin{subequations}
\label{relations_among_Sigmas}
\begin{align}
\Sigma_i^{(b)}(w) - \Sigma_i^{(c)}(w) & = 0 \qquad \qquad i=1, 4, 5, 6, 7 \;, \\
\Sigma_2^{(b)}(w) - \Sigma_2^{(c)}(w) & = \tilde{\Lambda} \, \Xi \;, \\
\Sigma_3^{(b)}(w) - \Sigma_3^{(c)}(w) & = - \tilde{\Lambda}' \, \Xi(w) \;.
\end{align}
\end{subequations}

The same procedure allows us to parametrize the matrix elements relevant for the $1 / m_Q^2$ terms in \eqref{current}:
\begin{subequations}
\begin{align}
\label{eqderRR}
\braket{ M'(v') | \bar{\psi}'_+ \, \Gamma \, i \, \Dright_\alpha \, i \, \Dright_\beta \, \psi_+ | M(v) } & = -\tr \big[ \Omega_{\mu\alpha\beta}^{(b)} \, \overline{\mathcal{M}}'^\mu \, \Gamma \, \mathcal{M} \big] \; \\
\label{eqderLL}
\braket{ M'(v') | \bar{\psi}'_+ \, i \, \Dleft_\alpha \, i \, \Dleft_\beta \, \Gamma \, \psi_+ | M(v) } & =- \tr \big[ \Omega_{\mu\alpha\beta}^{(c)} \, \overline{\mathcal{M}}'^\mu \, \Gamma \, \mathcal{M} \big] \;.
\end{align}
\end{subequations}
 $\Omega_{\mu\alpha\beta}^{(Q)}$ have the general expression

\bea
\Omega_{\mu\alpha\beta}^{(Q)} & =& \Omega_1^{(Q)} \, g_{\mu\alpha} \, v_\beta + \Omega_2^{(Q)} \, g_{\alpha\beta} \, v_\mu + \Omega_3^{(Q)} \, g_{\beta\mu} \, v_\alpha + \Omega_4^{(Q)} \, g_{\mu\alpha} \, v'_\beta + \Omega_5^{(Q)} \, g_{\beta\mu} \, v'_\alpha + \Omega_6^{(Q)} \, g_{\mu\alpha} \, \gamma_\beta  \nn \\
&+& \Omega_7^{(Q)} \, g_{\alpha\beta} \, \gamma_\mu + \Omega_8^{(Q)} \, g_{\beta\mu} \, \gamma_\alpha + \Omega_9^{(Q)} \, v_\mu \, v_\alpha \, v_\beta + \Omega_{10}^{(Q)} \, v_\mu \, v_\alpha \, v'_\beta + \Omega_{11}^{(Q)} \, v_\mu \, v'_\alpha \, v_\beta  \nn \\
& +& \Omega_{12}^{(Q)} \, v_\mu \, v'_\alpha \, v'_\beta + \Omega_{13}^{(Q)} \, v_\mu \, v_\alpha \, \gamma_\beta + \Omega_{14}^{(Q)} \, v_\mu \, \gamma_\alpha \, v_\beta + \Omega_{15}^{(Q)} \, \gamma_\mu \, v_\alpha \, v_\beta + \Omega_{16}^{(Q)} \, v_\mu \, v'_\alpha \, \gamma_\beta   \label{Omegas}\\
& +& \Omega_{17}^{(Q)} \, \gamma_\mu \, v_\alpha \, v'_\beta + \Omega_{18}^{(Q)} \, v_\mu \, \gamma_\alpha \, v'_\beta + \Omega_{19}^{(Q)} \, \gamma_\mu \, v'_\alpha \, v_\beta + \Omega_{20}^{(Q)} \, \gamma_\mu \, v'_\alpha \, v'_\beta + \Omega_{21}^{(Q)} \, i \, \sigma_{\mu\alpha} \, v_\beta  \nn \\
& +& \Omega_{22}^{(Q)} \, i \, \sigma_{\alpha\beta} \, v_\mu + \Omega_{23}^{(Q)} \, i \, \sigma_{\beta\mu} \, v_\alpha + \Omega_{24}^{(Q)} \, i \, \sigma_{\mu\alpha} \, v'_\beta + \Omega_{25}^{(Q)} \, i \, \sigma_{\beta\mu} \, v'_\alpha \;. \nn
\eea
The terms proportional to $v'^\mu$ vanish.
The matrix element with two derivatives, one acting on $ \psi_+$, the other one on  $\bar{\psi}'_+$,  can be parametrized integrating by parts either Eq.~\eqref{eqderRR},
\begin{equation}
\label{eqderLRdx}
\begin{split}
\braket{ M'(v') | \bar{\psi}'_+ \, (-i \, \Dleft_\alpha) \, \Gamma \, i \, \Dright_\beta \,  \psi_+ | M(v) } & =   \big( \tilde{\Lambda} \, v_\alpha - \tilde{\Lambda}' \, v'_\alpha \big) \, \tr \big[ \Sigma_{\mu\beta}^{(b)} \, \overline{\mathcal{M}}'^\mu \, \Gamma \, \mathcal{M} \big] - \tr \big[ \Omega_{\mu\alpha\beta}^{(b)} \, \overline{\mathcal{M}}'^\mu \, \Gamma \, \mathcal{M} \big] ,
\end{split}
\end{equation}
or  Eq.~\eqref{eqderLL},
\begin{equation}
\label{eqderLRsx}
\begin{split}
\braket{ M'(v') | \bar{\psi}'_+ \,(- i \, \Dleft_\alpha) \, \Gamma \, i \, \Dright_\beta \, \psi_+ | M(v) } & =- \big( \tilde{\Lambda} \, v_\beta - \tilde{\Lambda}' \, v'_\beta \big) \, \tr \big[ \Sigma_{\mu\alpha}^{(c)} \, \overline{\mathcal{M}}'^\mu \, \Gamma \, \mathcal{M} \big] - \tr \big[ \Omega_{\mu\alpha\beta}^{(c)} \, \overline{\mathcal{M}}'^\mu \, \Gamma \, \mathcal{M} \big] .
\end{split}
\end{equation}
This gives 
\be
\label{relations_among_psi}
\Omega_{\mu\alpha\beta}^{(b)}-\Omega_{\mu\alpha\beta}^{(c)}=\big( \tilde{\Lambda} \, v_\alpha - \tilde{\Lambda}' \, v'_\alpha \big) \, \Sigma_{\mu\beta}^{(b)}+\big( \tilde{\Lambda} \, v_\beta - \tilde{\Lambda}' \, v'_\beta \big) \,  \Sigma_{\mu\alpha}^{(c)} .
\ee

To express the form factors in the effective theory we have also to take into account the corrections from the expansion of the states.  They can be parametrized as 
\bea 
&&\braket{ M'(v') | i \, \int \diff^4 x \, \tprod \left[ J_0(0), \lagr_1(x) \right] | M(v) } = \hskip 2cm \nn \\
& &- \frac{1}{4 \, m_b} \, \underbrace{\left( - \frac{i}{2} \right) \, \tr \big[ \Upsilon_{2\mu\alpha\beta}^{(b)} \, \overline{\mathcal{M}}'^\mu \, \Gamma \, P_+ \, \sigma^{\alpha\beta} \, \mathcal{M} \big]}_{G^{(b)}} - \frac{1}{2 \, m_b^2} \, \underbrace{\tr \big[ \Upsilon_{1\mu}^{(b)} \, \overline{\mathcal{M}}'^\mu \, \Gamma \, \mathcal{M} \big]}_{K^{(b)}} \;,
\label{NL1} \\
&& \braket{ M'(v') | i \, \int \diff^4 x \, \tprod \left[ J_0(0), \lagr'_1(x) \right] | M(v) } = \nn \\
&& - \frac{1}{4 \, m_c} \, \underbrace{\left( - \frac{i}{2} \right) \, \tr \big[ \Upsilon_{2\mu\alpha\beta}^{(c)} \, \overline{\mathcal{M}}'^\mu \, \sigma^{\alpha\beta} \, P'_+ \, \Gamma \, \mathcal{M} \big]}_{G^{(c)}} - \frac{1}{2 \, m_c^2} \, \underbrace{\tr \big[ \Upsilon_{1\mu}^{(c)} \, \overline{\mathcal{M}}'^\mu \, \Gamma \, \mathcal{M} \big]}_{K^{(c)}} \;, \label{NL2}
\eea
where $\lagr_1, \lagr'_1$ are given in \eqref{eq:l1} with $m_Q\to m_b, m_c$, respectively.
The functions $\Upsilon^{(Q)}$  have the general expression
\bea
\Upsilon_{1\mu}^{(Q)} & =& \Upsilon_{1A}^{(Q)} \, v_\mu + \Upsilon_{1B}^{(Q)} \, \gamma_\mu \;, \label{upsilon1} \\
\Upsilon_{2\mu\alpha\beta}^{(Q)} &=& \Upsilon_{2A}^{(Q)} \, ( g_{\mu\alpha} \, v_\beta - g_{\mu \beta} \, v_\alpha) + \Upsilon_{2B}^{(Q)} \, (g_{\mu\alpha} \, v'_\beta - g_{\mu \beta} \, v'_\alpha) +  
\Upsilon_{2C}^{(Q)} \, (g_{\mu\alpha} \, \gamma_\beta - g_{\mu \beta} \, \gamma_\alpha)  \nn \\
&+& \Upsilon_{2D}^{(Q)} 
\, v_\mu \, ( v_\alpha \, v'_\beta - v'_\alpha \, v_\beta) +  
\Upsilon_{2E}^{(Q)} \,v_\mu \, ( v_\alpha \, \gamma_\beta - \gamma_\alpha \, v_\beta) +  \Upsilon_{2F}^{(Q)} \, v_\mu \, ( v'_\alpha \, 
\gamma_\beta - \gamma_\alpha \, v'_\beta)   \label{upsilon2} \\
&+& \Upsilon_{2G}^{(Q)} \, \gamma_\mu 
\, (v_\alpha \, v'_\beta - v'_\alpha \, v_\beta) + 
\Upsilon_{2H}^{(Q)} \, i \, ( \sigma_{\mu\alpha} \, v_\beta - \sigma_{\mu \beta} \, v_\alpha ) + \Upsilon_{2I}^{(Q)} \,i \, ( \sigma_{\mu\alpha} \, v'_\beta - \sigma_{\mu \beta} \, v'_\alpha ) 
\nn \\
&+& \Upsilon_{2J}^{(Q)}
\, i \, v_\mu \,  \sigma_{\alpha \beta}  \;.  \nn
\eea
Since the terms proportional to $v_\alpha$ or $v_\beta$ do not contribute to $G^{(b)}$, and those proportional to $v'_\alpha$ or $v'_\beta$ do not contribute to $G^{(c)}$,   we have:
 $\Upsilon_{2D}^{(Q)}=\Upsilon_{2G}^{(Q)}=0$ (for  $Q=c,b$),  $\Upsilon_{2B}^{(c)} =\Upsilon_{2F}^{(c)}=\Upsilon_{2I}^{(c)}=0$ and $\Upsilon_{2A}^{(b)} =\Upsilon_{2E}^{(b)}=\Upsilon_{2H}^{(b)}=0$.

With this set of relations,  the form factors in Eqs.~(\ref{ff:chic0})-(\ref{ff:hc})  are expressed in terms of  universal functions up to ${\cal O}(1/m^2_Q)$. The formulae are collected in Appendix \ref{app:universalff}.

\section{Relations  among form factors }\label{sec4}
At the leading order all form factors in $B_c \to \chi_{c0,1,2}, h_c$ matrix elements are expressed near zero recoil in terms of the single function $\Xi(w)$,
see  Appendix \ref{app:universalff}. 
Increasing the order of the expansion, relations among the form factors can be worked out exploiting the results in Appendix \ref{app:relations}.

The expansion in the relative quark velocity $\tilde v$ involves a large number of universal functions, which renders difficult  the derivations of the relations among form factors. It is interesting to consider the various orders in $1/m_Q$. In particular, 
at ${\cal O}(1/m_Q)$ there are relations among form factors in the same decay mode, in pairs of decay modes and in more than two modes, near the zero-recoil point.
 \\

\noi{\bf i)  Relations among the form factors in the same channel}

For the  $B_c \to \chi_{ci}$ (i=0,1,2) and  $B_c \to h_{c}$   form factors we have:
\begin{itemize}
\item $B_c \to \chi_{c0}$: 
\be
g_T(w) = - \frac{1}{w+1} \big[ 2 g_-(w) + g_P(w) \big] \, . \label{eq:relchic0}
\ee

\item $B_c \to \chi_{c1}$: 
\bea
g_{T_2} (w) &=& - \frac{1}{2} \big[ g_{V_1}(w) - (1+w) g_A(w) \big] \\
g_{T_3}(w) &=& - \frac{1}{2 ( w - 1 )} \big[ g_{V_1}(w) + 4 g_{V_2}(w) \big] + \frac{1}{2} g_A(w) + \frac{1}{w - 1} \big[ g_S(w) + g_{T_1}(w) \big]  \quad \label{eq:relchic1}
\eea
with  the condition 
\be
-\frac{1}{2} \big[ g_{V_1}(1) + 4 g_{V_2}(1) \big]  +  g_S(1) + g_{T_1}(1) = 0  \, .
\ee

\item $B_c \to \chi_{c2}$: 
\bea
k_{T_1}(w) &=& - w k_V(w) + k_{A_2}(w) + w k_{A_3}(w) + k_P(w) \\
k_{T_2}(w) &=& k_V(w) - k_{A_1}(w) - k_{A_2}(w) - w k_{A_3}(w) - k_P(w)  \label{eq:relchic2} \\
k_{T_3}(w) &=& - k_V(w) + k_{A_3}(w)  \, . 
\eea

\item $B_c \to h_c$:
\bea
f_{T_2}(w) &=& \frac{1}{2} \big[ f_{V_1}(w) + (1+w) f_A(w) \big] \\
f_{T_3}(w) &=& \frac{1}{2 ( w - 1 )} \big[ f_{V_1}(w) + 4 f_{V_2}(w) \big] + \frac{1}{2} f_A(w) - \frac{1}{w - 1} \big[ f_S(w) - f_{T_1}(w) \big]  \quad \label{eq:relhc}
\eea
with the condition
\be
 \frac{1}{2 } \big[ f_{V_1}(1) + 4 f_{V_2}(1) \big] -  \big[ f_S(1) - f_{T_1}(1) \big] = 0 .
\ee
\end{itemize}

\noi{\bf ii)  Relations among form factors of pairs of decay channels}

We have:

\begin{itemize}
\item $B_c \to \chi_{c0}$ and $B_c \to \chi_{c1}$: 
\bea
&&
( w+1 ) g_+(w) - ( w - 1 ) g_-(w) + g_P(w) = \nn \\
&&  \frac{w + 1}{\sqrt{6} } \Big\{ 2 g_{V_1}(w) + ( w + 1 ) g_{V_2}(w) - ( w - 1 ) \big[ g_{V_3}(w) + g_A(w) \big] - g_S(w) + 2 g_{T_1}(w) \Big\} . \quad \nn \\
\eea

\item $B_c \to h_c$ and $B_c \to \chi_{c1}$: 
\be
f_{V_1}(w) + ( w - 1 ) f_A(w) - 2 f_{T_1}(w) = 
  \sqrt{2} \Big\{ g_{V_1}(w) + ( w + 1 ) g_{V_2}(w) - ( w - 1 ) g_{V_3}(w) - g_S(w) \Big\}  
\ee
\bea
&&
3 f_{V_1}(w) + 2 ( w + 1 ) f_{V_2}(w) - ( w - 1 ) \big[ 2 f_{V_3}(w) - f_A(w) \big] - 2 \big[ f_S(w) + f_{T_1}(w) \big] = \qquad  \nn \\
&&  \sqrt{2} \Big\{ g_{V_1}(w) - ( w - 1 ) g_A(w) + 2 g_{T_1}(w) \Big\} \,\, .
\eea

\end{itemize}

\section{Phenomenology }\label{sec:discussion}
We now discuss some consequences of the relations found in the previous Sections. We use 
$m_{B_c}=6274.47\pm0.28\pm0.17$ MeV,  $\tau_{B_c}=0.510\pm0.009$ ps, $m_{\chi_{c0}(1P)}=3414.71\pm0.30$ MeV,  
$m_{\chi_{c1}(1P)}=3510.67\pm0.05$ MeV,   $m_{\chi_{c2}(1P)}=3556.17\pm0.07$ MeV and  $m_{h_{c}(1P)}=3525.38\pm0.11$ MeV \cite{Workman:2022ynf}.  
We first focus on the LO  relations, then on  NLO, mainly considering the Standard Model.
The form factors are expressed in terms of universal functions in a selected kinematic range, close to the zero recoil point $w=1$. In the numerical analyses we  extrapolate the relations to the full kinematic range in the various channels.  The range is  not wide ($w_{max}\sim 1.16-1.09$ for $B_c \to \chi_{c0}$, $B_c \to \chi_{c2}$ and light leptons or $\tau$, $w_{max}\sim 1.11-1.05$ for  $2P$ charmonia).  This allows  to get  information on the results in the full kinematical range.  In general,  the extrapolation can be  constrained by making use of dispersive matrices,   which allow to reconstruct the form factors knowing their values in few kinematical points \cite{Martinelli:2021frl}: the application of such methods is beyond the purposes of the present study.

\begin{figure}[!tb]
\begin{center}
\includegraphics[width = 0.405\textwidth]{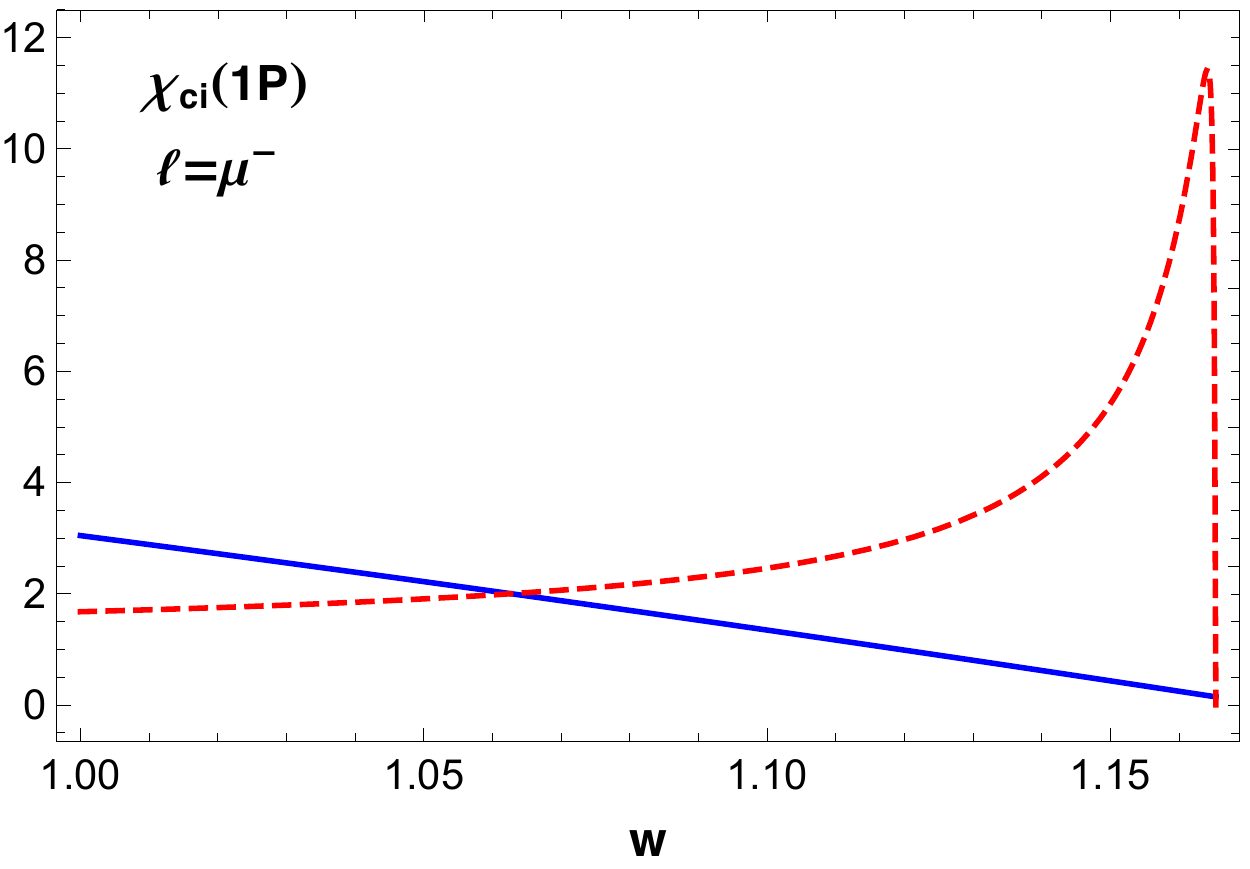}\hskip 0.4cm \includegraphics[width = 0.4\textwidth]{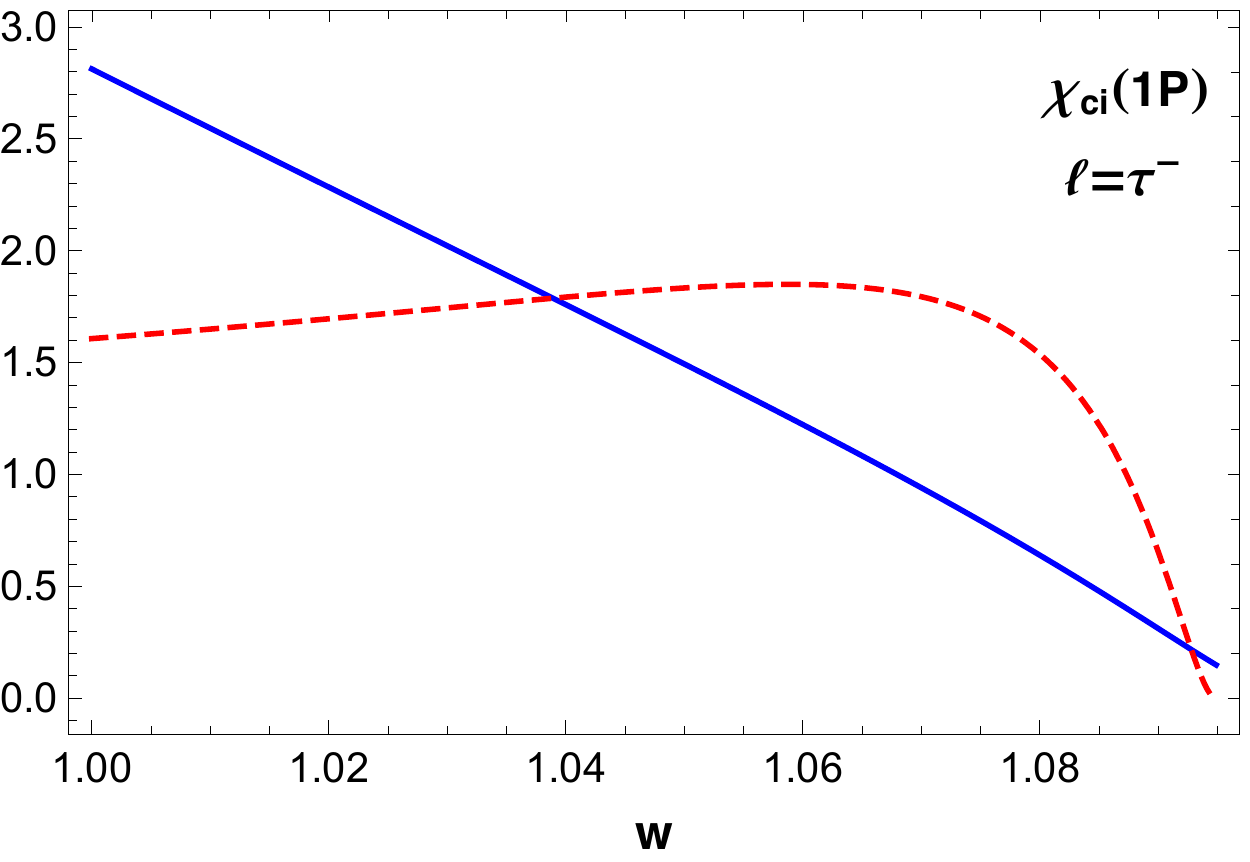}\\
\includegraphics[width = 0.4\textwidth]{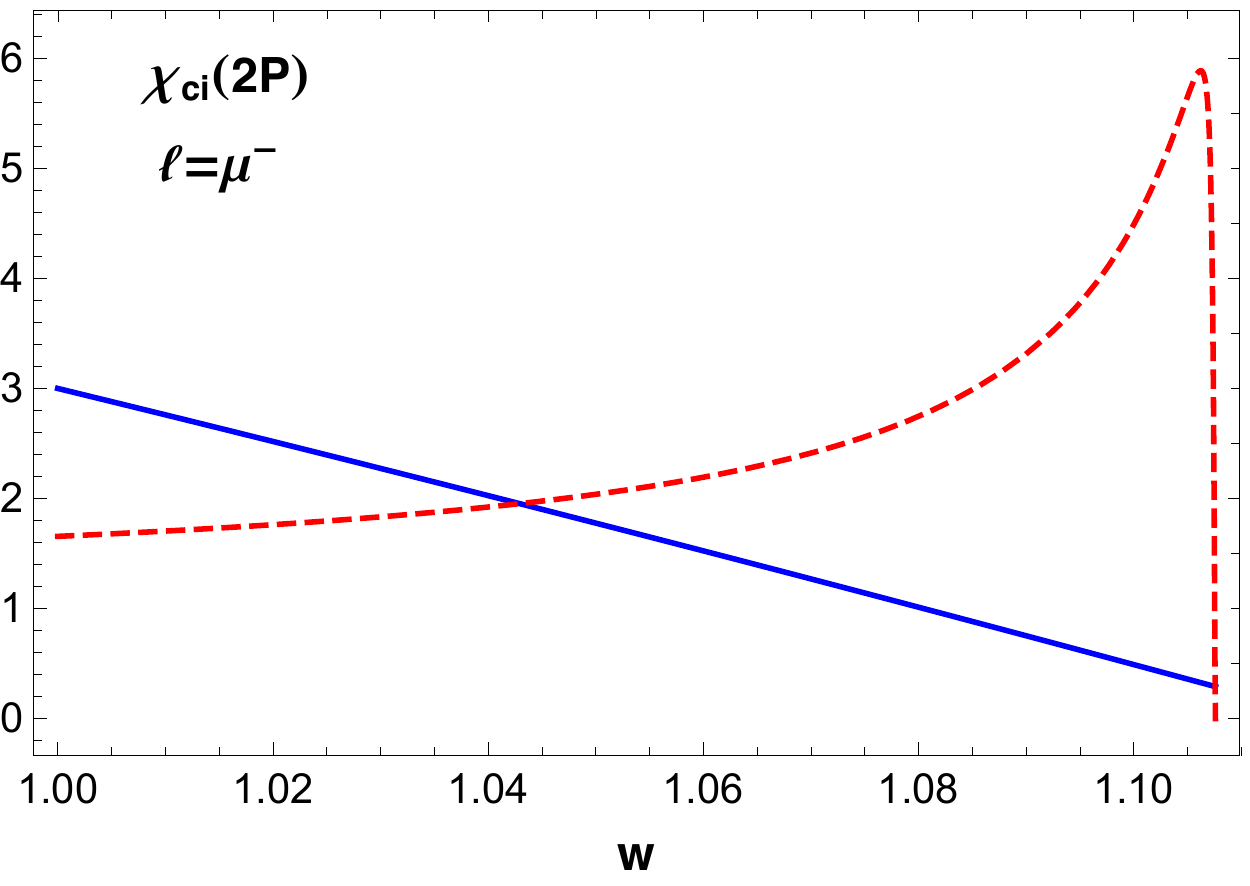}\hskip 0.4cm \includegraphics[width = 0.405\textwidth]{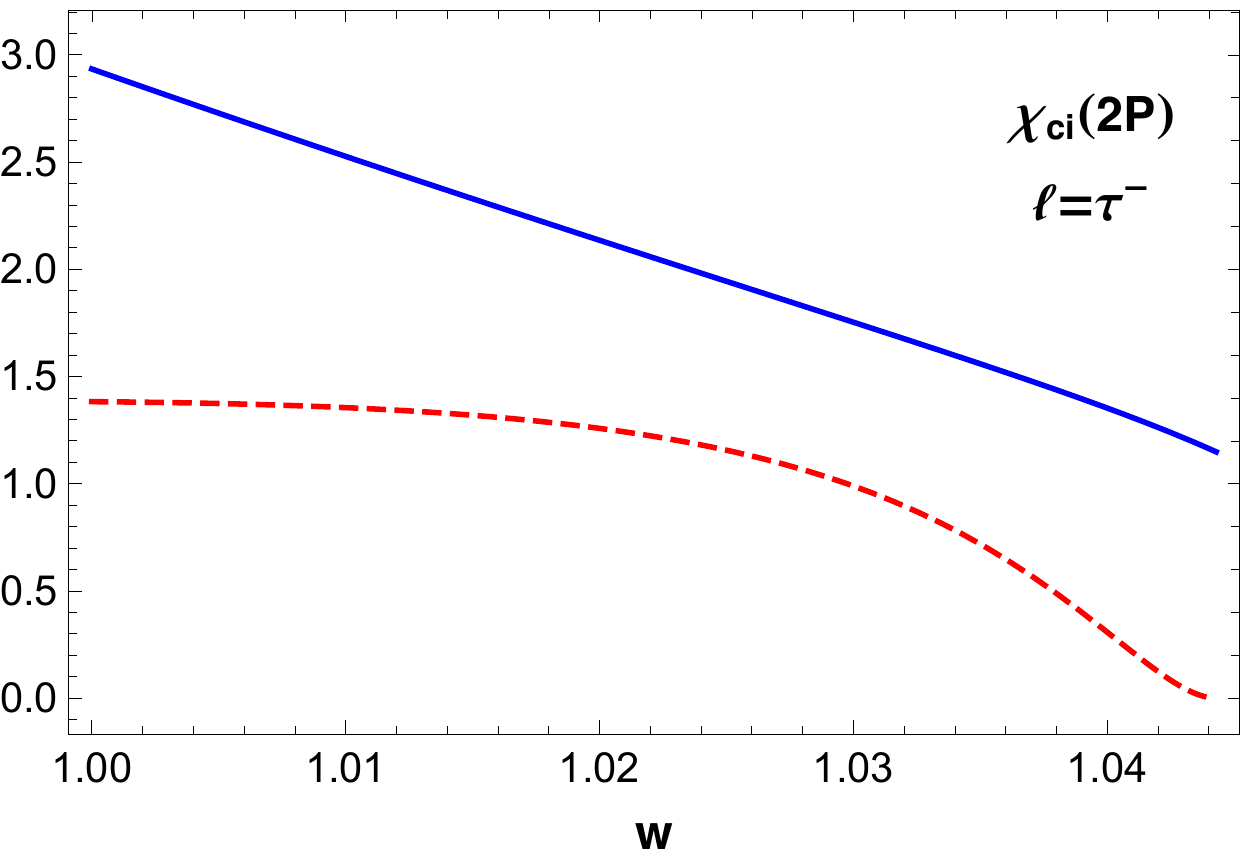}
    \caption{\small  Ratios of decay distributions  $\frac{d \Gamma(B_c \to \chi_{c1} \ell \bar \nu)/dw}{d \Gamma(B_c \to \chi_{c0} \ell \bar \nu)/dw}$ (continuous blue line) and 
    $\frac{d \Gamma(B_c \to \chi_{c2} \ell \bar \nu)/dw}{d \Gamma(B_c \to \chi_{c1} \ell \bar \nu)/dw}$  (dashed red line) in the Standard Model, in the case $\ell=\mu$ (left) and $\ell=\tau$ (right) for the $1P$ (top row) and $2P$ final charmonia (bottom row), with the meson masses  quoted in the text. The LO relations among form factors are extrapolated to the full kinematical range. }\label{fig1}
\end{center}
\end{figure}

The LO relations obtained from the formulae in appendix \ref{app:universalff} connect all  form factors of $B_c$ transitions to the $P-$wave charmonium 4-plet to the single function  $\Xi(w)$, which is different for the  $B_c \to 1P$ and the  $B_c \to 2P$ modes. In ratios of decay distributions, for the same value of $w$ the form factor dependence cancels out. The ratios are depicted in fig.~\ref{fig1}
both for $1P$ and the $2P$ channels, in the  range of $w$ common to all modes. For the $2P$ mesons we use the masses 
$m_{\chi_{c0}}(2P)=3860$ MeV and 
$m_{\chi_{c2}}(2P)=3930$ MeV \cite{Workman:2022ynf}, even though another $J^{PC}=0^{++}$ state is also reported by the Particle Data Group, $\chi_{c0}(3915)$ \cite{Workman:2022ynf}. 
%
%
 %
 The results  in the figure can be understood considering the  connection  among the decay distributions, holding at this order  if the $\chi_{ci}$ mass differences are neglected:
 \be
2 \frac{d \Gamma}{dw}(B_c \to \chi_{c0} \ell {\bar \nu}_\ell)+\frac{d \Gamma}{dw}(B_c \to \chi_{c1} \ell {\bar \nu}_\ell)-\frac{d \Gamma}{dw}(B_c \to \chi_{c2} \ell {\bar \nu}_\ell)=0 . \label{eq:relwidth}
\ee
Notice that this relation holds  in the SM and also including the full set of NP operators in the generalized Hamiltonian \eqref{hamil} regardless of  the Wilson coefficients $\epsilon_i$.

\begin{figure}[bt]
\begin{center}
\includegraphics[width = 0.40\textwidth]{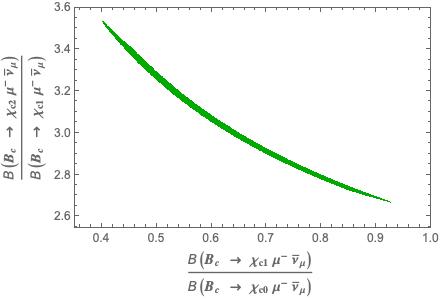}\hskip 0.4cm \includegraphics[width = 0.405\textwidth]{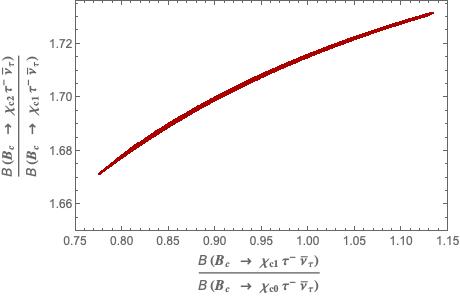}
\caption{\small   Correlations between the ratios of branching fractions $ \frac{{\cal B}(B_c \to \chi_{c2} \ell \bar \nu)}{{\cal B}(B_c \to \chi_{c1} \ell \bar \nu)}$ and 
 $ \frac{{\cal B}(B_c \to \chi_{c1} \ell \bar \nu)}{{\cal B}(B_c \to \chi_{c0} \ell \bar \nu)}$ 
for  $\ell=\mu$  (left) and  $\ell=\tau$ (right). The LO expression of the form factors is considered, with $\Xi(w)$ parametrized as in Eq.~\eqref{eq:xiexp}.  
 }\label{fig2}
\end{center}
\end{figure}

\begin{figure}[!bh]
\begin{center}
\includegraphics[width = 0.4\textwidth]{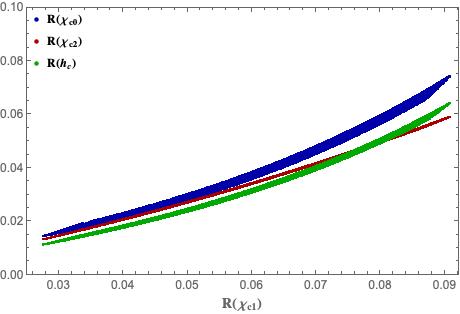}
    \caption{\small  Correlations between Lepton Flavour Universality ratios  $R(\chi_{c0,2})$ and $R(h_{c})$ with $R(\chi_{c1})$  in the SM. The LO expression of the form factors is considered, with $\Xi(w)$ parametrized in Eq.~\eqref{eq:xiexp}. }
    \label{fig3}
\end{center}
\end{figure}

\begin{figure}[t]
\begin{center}
\includegraphics[width = 0.405\textwidth]{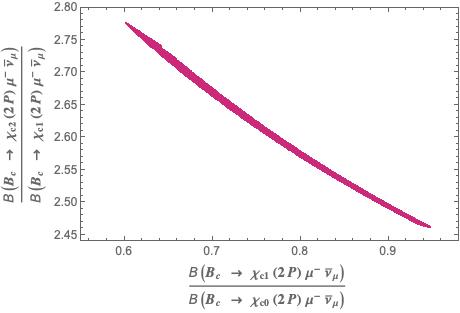}\hskip 0.4cm \includegraphics[width = 0.405\textwidth]{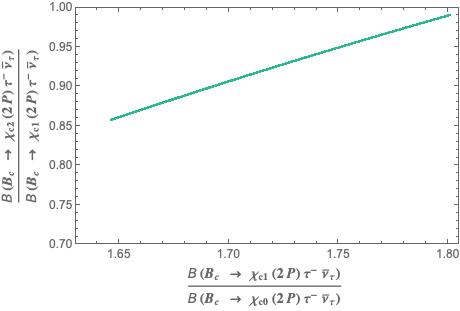}
\caption{\small   Correlation between ratios of branching fractions of $B_c$ decays to $2P$ charmonia and $\ell=\mu$ (left), $\ell=\tau$  (right), with the LO expression of the form factors  and  $\Xi(w)$  parametrized  in Eq.~\eqref{eq:xiexp}. }\label{fig4}
\end{center}
\end{figure}

\begin{figure}[!bh]
\begin{center}
\includegraphics[width = 0.4\textwidth]{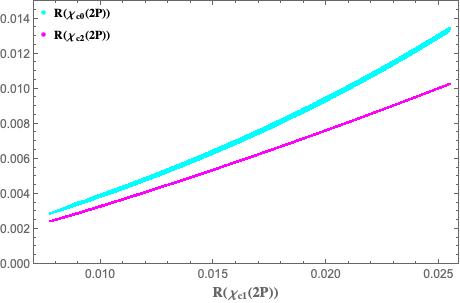}
    \caption{\small  Correlations between Lepton Flavour Universality ratios  $R(\chi_{c0,2}(2P))$  with $R(\chi_{c1}(2P))$  in the SM. The LO expression of the form factors is considered, with $\Xi(w)$ parametrized in Eq.~\eqref{eq:xiexp}. }
    \label{fig5}
\end{center}
\end{figure}

A simple parametrization of the function $\Xi(w)$ involves the intercept, slope and curvature at the zero recoil point,
\be
\Xi(w)=\Xi_0+\Xi_1  (w-1) + \Xi_2 (w-1)^2 \,\, . \label{eq:xiexp}
\ee
   The three parameters $\Xi_0,\,\Xi_1,\, \Xi_2$  can be constrained, namely $\Xi_0 \in [0.05,1]$, $\Xi_1 \in [-1,1]$ and  $\Xi_2 \in [-1,1]$, requiring that the measurement ${\cal B}(B_c^- \to \chi_{c0} \pi^-)=(2.4 \pm^{0.9}_{0.8})\times 10^{-5}$ \cite{Workman:2022ynf} is reproduced at 1$\sigma$ by naive factorization.
 Correlations between ratios of branching fractions are found varying  $\Xi_1/\Xi_0$ and $\Xi_2/\Xi_0$ in the selected regions. We choose  the same ranges also for  $2P$ excitations in the final state.  For the $B_c$ transitions to the $1P$ charmonium  a negative (positive) correlation is found between the ratios  $\dd \frac{\Gamma(B_c \to \chi_{c2} \ell \bar \nu_\ell)}{\Gamma(B_c \to \chi_{c1} \ell \bar \nu_\ell)}$ 
and  $\dd \frac{\Gamma(B_c \to \chi_{c1} \ell \bar \nu_\ell)}{\Gamma(B_c \to \chi_{c0} \ell \bar \nu_\ell)}$ for $\ell=\mu \, \, (\tau$),  as in fig.~\ref{fig2}.
There are positive correlations  also between the Lepton Flavour Universality ratios 
$R(C)=\Gamma(B_c \to C \tau \bar \nu_\tau)/\Gamma(B_c \to C \ell \bar \nu_\ell)$ which compare the decay rates to  $\tau$ and  to light leptons, see fig.~\ref{fig3}. 
%
%
%
At the same LO, the results for $B_c$ decays to the $2P$ charmonium are  shown in figs. \ref{fig1}, \ref{fig4} and \ref{fig5},  the behaviour expected for $\chi_{c1}(3872)$ if it corresponds to $\chi_{c1}(2P)$.

The above results  are independent of the  function $\Xi$, a benefit of the expansion of the form factors dominated by the leading order. Including the NLO terms  the phenomenology is more involved: nevertheless, the expansion allows us to perform systematic analyses and improvements.
An important feature is that the obtained relations connect the form factors to the matrix elements of operators in the effective theory:  they should be verified in  explicit calculations, and represent a testing ground for the various computations based on QCD methods. This is the case of
the relations among the form factors in the same channel, such as \eqref{eq:relchic0} for  $J^{PC}=0^{++}$, and \eqref{eq:relchic1}, \eqref{eq:relchic2} for the $J^{PC}=1^{++}, 2^{++}$ modes. 

 At ${\cal O}(1/m_Q)$, once the  relations  in Appendix C are taken into account, the number of  independent structures is $13$ in terms of universal functions.   For $\chi_{c0},\,\chi_{c1},\,h_c$ the decay distributions in SM are related for $\frac{1}{\tilde \Gamma} \frac{d\Gamma}{dw} \Big|_{w \to 1}$, with $\tilde \Gamma$ in \eqref{eq:gammatilde}. The expressions only involve $\Sigma_{\chi_{c1},1}^{(b)}$ and $\Sigma_{\chi_{c1},1}^{(c)}$ defined in Appendix \ref{app:universalff}:
 \bea
 \lim_{w \to 1} \frac{1}{\tilde \Gamma} \frac{d\Gamma}{dw}(B_c \to \chi_{c0} \ell {\bar \nu}_\ell) &=&18 \, {\hat m}_\ell^2 (\epsilon_b+\epsilon_c)^2\Big[\Sigma_{\chi_{c1},1}^{(b)}(1)\Big]^2  \\
  \lim_{w \to 1} \frac{1}{\tilde \Gamma} \frac{d\Gamma}{dw}(B_c \to \chi_{c1} \ell {\bar \nu}_\ell) &=&12\Big[2 (1-r_1)^2+{\hat m}_\ell^2 \Big]\Big[\epsilon_b\Sigma_{\chi_{c1},1}^{(b)}(1)-\epsilon_c\Sigma_{\chi_{c1},1}^{(c)}(1)\Big]^2  \\
    \lim_{w \to 1} \frac{1}{\tilde \Gamma} \frac{d\Gamma}{dw}(B_c \to h_c \ell {\bar \nu}_\ell) &=&6\Big[2 (1-r_h)^2+{\hat m}_\ell^2\Big]\Big[(\epsilon_b-\epsilon_c)\Sigma_{\chi_{c1},1}^{(b)}(1)+2\epsilon_c\Sigma_{\chi_{c1},1}^{(c)}(1)\Big]^2  \,\,.
 \eea
Such relations  also hold for the decays into the $2P$ resonances, and can be used to compare the mode involving $\chi_{c1}(3872)$ to other modes in the $2P$ 4-plet. Admittedly, this is a difficult measurement.

As for the NP extension in Eq.~(\ref{hamil}) with   the pseudoscalar  and   tensor operators included,  the form factors parametrizing the scalar and tensor matrix elements  give  contributions related to the SM ones as in  Eq.~(\ref{eq:relchic0}), a  useful connection for phenomenological analyses.

Other improvements are possible if reliable results  are available even for a single form factor.  As shown  in \cite{Colangelo:2022lpy} using input from lattice NRQCD, details on a  form factor can be employed to get information on  universal functions, predicting  other form factors and  establishing connections among observables.

\section{Conclusions }\label{conc}

We have derived the expressions of the form factors of the semileptonic $B_c$ decays to the $P$-wave charmonium 4-plet as an expansion in the relative velocity of the heavy quarks in the charmonium and of $1/m_Q$. The expressions involve universal functions, independent of the specific channel, and allow to connect different modes. They are a testing ground for explicit calculations of the form factors and  can be used in studying LFU ratios and the effects of SM extensions. They are useful  in  analyzing the semileptonic $B_c$ decays to the $2P$ charmonia: the comparison of measurements of  $B_c$ transitions to $\chi_{c1}(3872)$  with  other  states in the $2P$ charmonium 4-plet  is a  tool to gain new information on the  nature of $X(3872)$.

 \vspace*{1.cm}
\noindent {\bf Acknowledgements.}
This study has been  carried out within the INFN project (Iniziativa Specifica) QFT-HEP.
\appendix
\numberwithin{equation}{section}
\section{Another parametrization of the $B_c \to \chi_{c\, 0,1,2}$  and $B_c \to h_c$ matrix elements}\label{app:ff}
For the sake of completeness we report another parametrization of the $B_c \to \chi_{c\, 0,1,2}$  and $B_c \to h_c$ matrix elements in terms of form factors, often used in the literature:
\bea
\langle S(p')|\bar{c}\gamma^\mu \gamma_5 b|{B_c} (p)\rangle
&=& f_0^{S}(q^2) \frac{m_{B_c}^2 - m_{S}^2}{q^2} q^{\mu} + f_+^{S}(q^2) \left( p^{\mu} + p'^{\mu} - \frac{m_{B_c}^2 - m_{S}^2}{q^2} q^{\mu} \right), \nn \\
\langle S(p')|\bar{c}\sigma^{\mu \nu} b|{B_c} (p)\rangle
&=& - \frac{2 f_T^{S}(q^2)}{m_{B_c} + m_{S}} \varepsilon^{\mu\nu\rho\sigma} p_{\rho} p'_{\sigma},  \label{ffS} \\
\langle S(p')|\bar{c}\sigma^{\mu \nu} \gamma_5 b|{B_c} (p)\rangle
&=& - i\frac{2 f_T^{S}(q^2)}{m_{B_c} + m_{S}} \left( p^{\mu} p'^{\nu}-p^{\nu} p'^{\mu} \right), \nn 
\eea
\bea
\langle A(p', \epsilon)|\bar{c}\gamma^\mu b|{B_c} (p)\rangle
&=& -i \Big[2 m_{A} A_0^{A}(q^2) \frac{\epsilon^* \cdot q}{q^2} q^{\mu} + (m_{B_c} + m_{A}) A_1^{A}(q^2) \left(\epsilon^{* \mu}-\frac{\epsilon^* \cdot q}{q^2} q^{\mu} \right) \nn \\
&-& A_2^{A}(q^2) \frac{\epsilon^* \cdot q}{m_{B_c}+m_{A}} \left(p^{\mu} + p'^{\mu}-\frac{m_{B_c}^2-m_{A}^2}{q^2} q^{\mu}\right) \Big], \nn \\
\langle A(p', \epsilon)|\bar{c}\gamma^\mu \gamma_5 b|{B_c} (p)\rangle
&=& \frac{2 V^{A}(q^2)}{m_{B_c} + m_{A}}\varepsilon^{\mu\nu\rho\sigma} \epsilon_{\nu}^* p_{\rho}' p_{\sigma}, \nn \\
\langle A(p',\epsilon)|\bar{c}\sigma^{\mu \nu} b|{B_c} (p)\rangle
&=& i\, T_0^{A}(q^2) \frac{\epsilon^* \cdot q} {(m_{B_c}+ m_A)^2} (p^\mu p^{\prime \nu}-p^\nu p^{\prime \mu})  \label{ffA}\\
 &+& i\, T_1^{A}(q^2) (p^\mu \epsilon^{* \nu}-\epsilon^{* \mu} p^\nu)  +i\,T_2^{A}(q^2)(p^{\prime \mu} \epsilon^{* \nu}-\epsilon^{* \mu} p^{\prime \nu}) , \nn \\
\langle A(p', \epsilon)|\bar{c}\sigma^{\mu \nu} \gamma_5 b|{B_c} (p)\rangle
&=& T_0^{A}(q^2) \frac{\epsilon^* \cdot q } {(m_{B_c}+ m_A)^2} \varepsilon^{\mu \nu \alpha \beta} p_\alpha p'_{\beta}+T_1^{A}(q^2) \varepsilon^{\mu \nu \alpha \beta} p_\alpha \epsilon^*_{\beta}   + T_2^{A}(q^2) \varepsilon^{\mu \nu \alpha \beta} p'_{\alpha} \epsilon^*_{\beta},\nn 
\eea
\bea
\langle T(p', \eta)|\bar{c}\gamma^\mu b|{B_c} (p)\rangle &=& \frac{2 V^{T}(q^2)}{m_{B_c} \left(m_{B_c} + m_{T}\right)} \varepsilon^{\mu\nu\rho\sigma} \eta_{\nu \alpha}^* p_{\rho}' p_{\sigma} q^{\alpha}, \nn \\
\langle T(p', \eta)|\bar{c}\gamma^\mu \gamma_5 b|{B_c} (p)\rangle
&=& \Big[2 m_{T} A_0^{T}(q^2) \frac{\eta^{* \alpha \beta} q_{\beta}}{q^2} q^{\mu} + (m_{B_c} + m_{T}) A_1^{T}(q^2) \left(\eta^{* \mu \alpha}-\frac{\eta^{* \alpha \beta} q_{\beta}}{q^2} q^{\mu}\right)  \nn \\ 
&-& A_2^{T}(q^2)\frac{\eta^{* \alpha \beta} q_{\beta}}{m_{B_c} + m_{T}} \left(p^{\mu} + p'^{\mu} -\frac{m_{B_c}^2-m_{T}^2}{q^2} q^{\mu}\right) \Big] \frac{-i q_{\alpha}}{m_{B_c}}, \nn \\
\langle T(p', \eta)|\bar{c} \sigma^{\mu\nu} b|{ B} (p)\rangle
&=& \Big[T_0^{T}(q^2) \frac{\eta^*_{\rho \tau} q^{\rho} } {(m_{B_c}+ m_T)^2} \varepsilon^{\mu \nu \alpha \beta} p'_\alpha p_{\beta}+T_1^{T}(q^2) \varepsilon^{\mu \nu \alpha \beta} \eta^*_{\alpha \tau} p_\beta \label{ffT} \\
&+& T_2^{T}(q^2) \varepsilon^{\mu \nu \alpha \beta} \eta^*_{\alpha \tau} p'_{\beta} \Big] \frac{q^{\tau}}{m_{B_c}} , \nn
\eea
\bea
\langle T(p', \eta)|\bar{c} \sigma^{\mu\nu}\gamma_5 b|{ B} (p)\rangle
&=& \Big[ T_0^{T}(q^2) \frac{\eta^{*\alpha \beta} q_{\alpha}} {(m_{B_c}+ m_T)^2} (p^{\mu} p^{\prime \nu} -p^\nu p^{\prime \mu}) \nn \\
 &+& T_1^{T}(q^2) (p^\mu \eta^{*\nu \beta} -\eta^{* \mu \beta} p^\nu)  + T_2^{T}(q^2)(p^{\prime \mu} \eta^{* \nu \beta} - \eta^{* \mu \beta } p^{\prime \nu}) \Big] \frac{-i q_{\beta}}{m_{B_c}} . \nn
\eea
 In Eqs.~\eqref{ffS}, \eqref{ffA} and \eqref{ffT}  we denote $S = \rchi_{c_0}$, $A = \rchi_{c_1}$,$h_c$, and $T = \rchi_{c_2}$.  The  conditions
\bea
f_0^S (0) &=& f_+^S(0) \nn \\
A_0^{A(T)}(0) &=& \frac{m_B + m_{A(T)}}{2 m_{A(T)}} A_1^{A(T)}(0) - \frac{m_B
- m_{A(T)}}{2 m_{A(T)}} A_2^{A(T)}(0)
\eea
remove the singularity at $q^2=0$.

\section{Form factors in terms of universal functions at $\mathcal{O}(1 / m^2_Q)$}\label{app:universalff}
\numberwithin{equation}{subsection}

In this Appendix we collect  the relations between the form factors in Eqs.~\eqref{ff:chic0}-\eqref{ff:hc} and those expressed in the effective theory at $\mathcal{O}(1 / m^2_Q)$.
We define $\epsilon_{b(c)}=\displaystyle\frac{1}{2m_{b(c)}}$ 
and 
$\Omega_{i}^{(Q), \text{mix}} = \Omega_{i}^{(b), \text{mix}} + \Omega_{i}^{(c), \text{mix}}$ (for $i=0,1,2$), with  $\Omega_{i}^{(b,c), \text{mix}} $ given below.

\subsection{$B_c \to \chi_{c0}$}

We  define:
\begin{align}
\Sigma_{\chi_{c0}}^{(b)} & = (2 + w) \Sigma_1^{(b)} + (w^2 - 1) \Sigma_3^{(b)} - 3 ( w - 1 ) [ \Sigma_4^{(b)} + \Sigma_6^{(b)} ] - ( w - 7 ) \Sigma_7^{(b)}
\;, \\
\Sigma_{\chi_{c0}}^{(c)} & = 3 \Sigma_1^{(c)} - (w^2 - 1) \Sigma_2^{(c)} - ( w - 1 ) [ \Sigma_4^{(c)} - 3 \Sigma_5^{(c)} ] + 6 \Sigma_7^{(c)}
\;, \\
\Omega_{\chi_{c0}}^{(b)} & = ( w + 2 ) [ \Omega_3^{(b)} + w \Omega_5^{(b)} - \Omega_8^{(b)} ] + ( w^2 - 1 ) [ \Omega_4^{(b)} + \Omega_{10}^{(b)} + w \Omega_{12}^{(b)} - \Omega_{18}^{(b)} ]  \notag \\
&  - 3 ( w - 1 ) [ \Omega_6^{(b)} + \Omega_{13}^{(b)} + w \Omega_{16}^{(b)} + \Omega_{17}^{(b)} + w \Omega_{20}^{(b)} + \Omega_{22}^{(b)} ] + ( w - 7 ) [ \Omega_{23}^{(b)} + w \Omega_{25}^{(b)} ]  \notag \\
&  - ( w - 1 ) ( w - 2 ) \Omega_{24}^{(b)}
\;, \\
\Omega_{\chi_{c0}}^{(c)} & = - 3 [ w \Omega_1^{(c)} + \Omega_4^{(c)} - \Omega_6^{(c)} + 2 w \Omega_{21}^{(c)} + 2 \Omega_{24}^{(c)} ] + ( w^2 - 1 ) [ w \Omega_9^{(c)} + \Omega_{10}^{c} - \Omega_{13}^{(c)} ]  \notag \\
&  + ( w - 1 ) [ w \Omega_{14}^{(c)} - 3 w \Omega_{15}^{(c)} - 3 \Omega_{17}^{(c)} + \Omega_{18}^{(c)} + \Omega_{22}^{(c)} + 3 \Omega_{23}^{(c)} ]
\;, \\
\Omega_{\chi_{c0}}^{(Q), \text{mix}} & = ( w + 1 ) ( w - 2 ) \Omega_2^{(Q)} + ( w + 1 ) ( w + 2 ) \Omega_3^{(Q)} - 3 ( w + 1 ) [ \Omega_4^{(Q)} + 2 \Omega_{24}^{Q} ] + 9 \Omega_6^{(Q)}  \notag \\
&  - ( w - 2 ) [ 3 \Omega_7^{(Q)} - \Omega_8^{(Q)} ] + ( w - 1 ) ( w + 1 )^2 \Omega_{10}^{(Q)} - ( w^2 - 1 ) [ 3 \Omega_{13}^{(Q)} + 3 \Omega_{17}^{(Q)} - \Omega_{18}^{(Q)} ]  \notag \\
&  - ( w + 1 )^2 \Omega_{22}^{(Q)} + ( w + 1 ) ( w - 7 ) \Omega_{23}^{(Q)}
\;, \\
\Upsilon_{2, \chi_{c0}}^{(b)} & = ( w + 1 ) [ 2 \Upsilon_{2B}^{(b)} + 2 ( w - 1 ) \Upsilon_{2F}^{(b)} + 4 \Upsilon_{2I}^{(b)} + 3 \Upsilon_{2J}^{(b)} ] + 2 ( w - 2 ) \Upsilon_{2C}^{(b)}
\;, \\
\Upsilon_{2, \chi_{c0}}^{(c)} & = ( w + 1 ) [ 2 \Upsilon_{2A}^{(c)} + 2 ( w - 1 ) \Upsilon_{2E}^{(c)} + 4 \Upsilon_{2H}^{(c)} - \Upsilon_{2J}^{(c)} ] - 6 \Upsilon_{2C}^{(c)}
\;.
\end{align}
%
Using these definitions we obtain the expressions of the form factors in terms of universal functions:
\begin{align}
g_+ & = - \frac{1}{\sqrt{3}} \big[ \epsilon_b \Sigma_{\chi_{c0}}^{(b)} + \epsilon_c \Sigma_{\chi_{c0}}^{(c)} \big] + \frac{1}{\sqrt{3}} \big[ \epsilon_b^2 \Omega_{\chi_{c0}}^{(b)} - \epsilon_c^2 \Omega_{\chi_{c0}}^{(c)} \big]
\;, \\
g_- & = \frac{w + 1}{\sqrt{3}} \Xi + \frac{1}{2 \sqrt{3}} \big[ \epsilon_b \Upsilon_{2, \chi_{c0}}^{(b)} + \epsilon_c \Upsilon_{2, \chi_{c0}}^{(c)} \big] \notag \\
 &+ \frac{2}{\sqrt{3}} \Big[ \epsilon_b^2 [ ( w + 1 ) \Upsilon_{1A}^{(b)} - 3 \Upsilon_{1B}^{(b)} ] + \epsilon_c^2 [ ( w + 1 ) \Upsilon_{1A}^{(c)} - 3 \Upsilon_{1B}^{(c)} ] \Big]\notag\\
&+ \frac{1}{2 \sqrt{3}} \epsilon_b \epsilon_c \big[ ( w + 1 )( \tilde{\Lambda} \Sigma_{\chi_{c0}}^{(b)} - \tilde{\Lambda}' \Sigma_{\chi_{c0}}^{(c)} ) -\, \Omega_{\chi_{c0}}^{(Q), \text{mix}} \big]  
\;, \\
g_P & = \frac{w^2 - 1}{\sqrt{3}} \Xi - \frac{w + 1}{\sqrt{3}} \big[ \epsilon_b \Sigma_{\chi_{c0}}^{(b)} - \epsilon_c \Sigma_{\chi_{c0}}^{(c)} \big] + \frac{w - 1}{2 \sqrt{3}} \big[ \epsilon_b \Upsilon_{2, \chi_{c0}}^{(b)} + \epsilon_c \Upsilon_{2, \chi_{c0}}^{(c)} \big]  \notag \\&  + \frac{2 (w - 1)}{\sqrt{3}} \Big[ \epsilon_b^2 [ ( w + 1 ) \Upsilon_{1A}^{(b)} - 3 \Upsilon_{1B}^{(b)} ] + \epsilon_c^2 [ ( w + 1 ) \Upsilon_{1A}^{(c)} - 3 \Upsilon_{1B}^{(c)} ] \Big] \notag \\
&+ \frac{w + 1}{\sqrt{3}} \big[ \epsilon_b^2 \Omega_{\chi_{c0}}^{(b)} + \epsilon_c^2 \Omega_{\chi_{c0}}^{(c)} \big] - \frac{w - 1}{2 \sqrt{3}} \epsilon_b \epsilon_c \big[ ( w + 1 )( \tilde{\Lambda} \Sigma_{\chi_{c0}}^{(b)} - \tilde{\Lambda}' \Sigma_{\chi_{c0}}^{(c)} ) -\, \Omega_{\chi_{c0}}^{(Q), \text{mix}} \big]  
\;, \\
g_T & = - \frac{w + 1}{\sqrt{3}} \Xi 
+ \frac{1}{\sqrt{3}} \big[ \epsilon_b \Sigma_{\chi_{c0}}^{(b)} - \epsilon_c \Sigma_{\chi_{c0}}^{(c)} \big] 
- \frac{1}{2 \sqrt{3}} \big[ \epsilon_b \Upsilon_{2, \chi_{c0}}^{(b)} + \epsilon_c \Upsilon_{2, \chi_{c0}}^{(c)} \big]  \notag \\
&  - \frac{2}{\sqrt{3}} \Big[ \epsilon_b^2 [ ( w + 1 ) \Upsilon_{1A}^{(b)} - 3 \Upsilon_{1B}^{(b)} ] + \epsilon_c^2 [( w + 1 ) \Upsilon_{1A}^{(c)} - 3 \Upsilon_{1B}^{(c)} ] \Big] \notag \\ &
- \frac{1}{\sqrt{3}} \big[ \epsilon_b^2 \Omega_{\chi_{c0}}^{(b)} + \epsilon_c^2 \Omega_{\chi_{c0}}^{(c)} \big] + \frac{1}{2 \sqrt{3}} \epsilon_b \epsilon_c \big[ ( w + 1 )( \tilde{\Lambda} \Sigma_{\chi_{c0}}^{(b)} - \tilde{\Lambda}' \Sigma_{\chi_{c0}}^{(c)} ) - \, \Omega_{\chi_{c0}}^{(Q), \text{mix}} \big] 
\;.
\end{align}

\subsection{$B_c \to \chi_{c1}$}

We define:
\begin{align}
\Sigma_{\chi_{c1}, 1}^{(b)} & = \Sigma_1^{(b)} - ( w - 1 ) \Sigma_6^{(b)} + 2 \Sigma_7^{(b)}
\;, \\
\Sigma_{\chi_{c1}, 2}^{(b)} & = \Sigma_1^{(b)} + ( w + 1 ) \Sigma_3^{(b)} - 3 \Sigma_4^{(b)} - \Sigma_7^{(b)}
\;, \\
\Sigma_{\chi_{c1}, 1}^{(c)} & = \Sigma_1^{(c)} - ( w - 1 ) \Sigma_5^{(c)} \;, \\
\Sigma_{\chi_{c1}, 2}^{(c)} & = ( w + 1 ) \Sigma_2^{(c)} - \Sigma_4^{(c)} - 4 \Sigma_5^{(c)}
\;, \\
\Omega_{\chi_{c1}, 1}^{(b)} & = - \Omega_3^{(b)} - w \Omega_5^{(b)} + \Omega_8^{(b)} + ( w - 1 ) [ \Omega_{17}^{(b)} + w \Omega_{20}^{(b)} - \Omega_{24}^{(b)} ] + 2 \, \Omega_{23}^{(b)} + 2 w \Omega_{25}^{(b)}
\;, \\
\Omega_{\chi_{c1}, 2}^{(b)} & = \Omega_3^{(b)} + ( w + 1 ) [ \Omega_{4}^{(b)} + \Omega_{10}^{(b)} + w \Omega_{12}^{(b)} - \Omega_{18}^{(b)} - \Omega_{24}^{(b)} ] \notag \\
&+ w \Omega_5^{(b)} - 3 \Omega_{6}^{(b)} - \Omega_8^{(b)} - 3 \Omega_{13}^{(b)}   - 3 w \Omega_{16}^{(b)} - 3 \Omega_{22}^{(b)} + \Omega_{23}^{(b)} + w \Omega_{25}^{(b)}
\;, \\
\Omega_{\chi_{c1}, 1}^{(c)} & = w \Omega_{1}^{(c)} + \Omega_4^{(c)} - \Omega_6^{(c)} - ( w - 1 ) [ w \Omega_{15}^{(c)}  +\Omega_{17}^{(c)} - \Omega_{23}^{(c)} ]
\;, \\
\Omega_{\chi_{c1}, 2}^{(c)} & = - ( w + 1 ) [ w \Omega_{9}^{(c)} + \Omega_{10}^{(c)} - \Omega_{13}^{(c)} ] + w [ \Omega_{14}^{(c)} + 4 \Omega_{15}^{(c)} ] \notag \\
&+ 4 \Omega_{17}^{(c)} + \Omega_{18}^{(c)} + \Omega_{22}^{(c)} - 4 \Omega_{23}^{(c)} 
\;, \\
\Omega_{\chi_{c1}, 1}^{(Q), \text{mix}} & = ( w + 1 ) [ \Omega_{3}^{(Q)} + \Omega_{4}^{(Q)} - 2 \Omega_{23}^{(Q)} ] - 3 \Omega_{6}^{(Q)} - ( w + 2 ) \Omega_{7}^{(Q)} - \Omega_{8}^{(Q)} - ( w^2 - 1 ) \Omega_{17}^{Q} 
\;, \\
\Omega_{\chi_{c1}, 2}^{(Q), \text{mix}} & = w \Omega_{2}^{(Q)} + ( w + 3 ) \Omega_{3}^{(Q)} - 4 \Omega_{7}^{(Q)} - \Omega_{8}^{(Q)} + ( w^2 - 1 ) \Omega_{10}^{(Q)}  \notag \\
&  - ( w - 1 ) [ 3 \Omega_{13}^{(Q)} + 4 \Omega_{17}^{(Q)} + \Omega_{18}^{(Q)} ] - ( w - 3 ) \Omega_{22}^{(Q)} + ( w - 9 ) \Omega_{23}^{(Q)}
\;, \\
\Upsilon_{2, \chi_{c1}, 1}^{(b)} & =(1+w) [ \Upsilon_{2B}^{(b)} + \Upsilon_{2I}^{(b)} ] - 2 \Upsilon_{2C}^{(b)} 
\;, \\
\Upsilon_{2, \chi_{c1}, 2}^{(b)} & = \Upsilon_{2B}^{(b)} - 2 \Upsilon_{2C}^{(b)} - 2 (w - 1) \Upsilon_{2F}^{(b)} - \Upsilon_{2I}^{(b)} - 3 \Upsilon_{2J}^{(b)}
\;, \\
\Upsilon_{2, \chi_{c1}, 1}^{(c)} & = \Upsilon_{2A}^{(c)} + \Upsilon_{2H}^{(c)}
\;, \\
\Upsilon_{2, \chi_{c1}, 2}^{(c)} & =3  \Upsilon_{2A}^{(c)} +  \Upsilon_{2H}^{(c)} -  \Upsilon_{2J}^{(c)}
\;.
\end{align}
%
With these definitions,  we express the form factors in terms of universal functions:
\begin{align}
g_{V_1} & = \frac{w^2 - 1}{\sqrt{2}} \Xi - \frac{w + 1}{\sqrt{2}} \Big[ \epsilon_b [ 2 \Sigma_{\chi_{c1}, 1}^{(b)} + ( w - 1 ) \Sigma_{\chi_{c1}, 2}^{(b)} ] - \epsilon_c [ 2 \Sigma_{\chi_{c1}, 1}^{(c)} - ( w - 1 ) \Sigma_{\chi_{c1}, 2}^{(c)} ] \Big]  \notag \\
&  + \frac{w - 1}{2 \sqrt{2}} \Big[ \epsilon_b [ 2 \Upsilon_{2, \chi_{c1}, 1} ^{(b)} - ( w + 1 ) \Upsilon_{2, \chi_{c1}, 2}^{(b)} ] - \epsilon_c ( w + 1 ) [ 2 \Upsilon_{2, \chi_{c1}, 1} ^{(c)} - \Upsilon_{2, \chi_{c1}, 2}^{(c)} ] \Big]  \notag \\
& + \sqrt{2} (w - 1) \Big[ \epsilon_b^2 [ (w + 1) \Upsilon_{1A} ^{(b)} - 2 \Upsilon_{1B} ^{(b)} )]+ \epsilon_c^2 [ (w + 1) \Upsilon_{1A} ^{(c)} - 2 \Upsilon_{1B} ^{(c)} ] \Big] \notag \\
& - \frac{w + 1}{\sqrt{2}} \Big[ \epsilon_b^2 [ 2 \Omega_{\chi_{c1}, 1}^{(b)} - ( w - 1 ) \Omega_{\chi_{c1}, 2}^{(b)} ] + \epsilon_c^2 [ 2 \Omega_{\chi_{c1}, 1}^{(c)} + ( w - 1 ) \Omega_{\chi_{c1}, 2}^{(c)} ] \Big]  \notag \\
&  - \frac{w - 1}{2 \sqrt{2}} \epsilon_b \epsilon_c \Big[ ( w + 1 ) \Big( \tilde{\Lambda} [ 2 \Sigma_{\chi_{c1}, 1}^{(b)} + ( w - 1 ) \Sigma_{\chi_{c1}, 2}^{(b)} ] - \tilde{\Lambda}' [ 2 \Sigma_{\chi_{c1}, 1}^{(c)} - ( w - 1 ) \Sigma_{\chi_{c1}, 2}^{(c)} ] \Big) \notag \\
& + [ 2 \Omega_{\chi_{c1}, 1}^{(Q), \text{mix}} - ( w + 1 ) \Omega_{\chi_{c1}, 2}^{(Q), \text{mix}} ] \Big] 
\;, \\
g_{V_2} & = - \frac{w - 1}{2 \sqrt{2}} \Xi + \frac{1}{2 \sqrt{2}} \Big[ \epsilon_b [ 2 \Sigma_{\chi_{c1}, 1}^{(b)} + ( w - 1 ) \Sigma_{\chi_{c1}, 2}^{(b)} ] - \epsilon_c [ 2 \Sigma_{\chi_{c1}, 1}^{(c)} - ( w - 1 ) \Sigma_{\chi_{c1}, 2}^{(c)} ] \Big]  \notag \\
& - \frac{1}{4 \sqrt{2}} \Big[ \epsilon_b [ 2 \Upsilon_{2, \chi_{c1}, 1} ^{(b)} - ( w - 1 ) \Upsilon_{2, \chi_{c1}, 2}^{(b)} ] - \epsilon_c [ 2 ( w + 1 ) \Upsilon_{2, \chi_{c1}, 1} ^{(c)} - ( w - 1 ) \Upsilon_{2, \chi_{c1}, 2}^{(c)} ] \Big]  \notag \\
&  - \frac{1}{\sqrt{2}} \Big[ \epsilon_b^2 [ (w - 1) \Upsilon_{1A} ^{(b)} - 2 \Upsilon_{1B} ^{(b)} ] + \epsilon_c^2 [ (w - 1) \Upsilon_{1A} ^{(c)} - 2 \Upsilon_{1B} ^{(c)} ] \Big]
\notag \\
& + \frac{1}{2 \sqrt{2}} \Big[ \epsilon_b^2 [ 2 \Omega_{\chi_{c1}, 1}^{(b)} - ( w - 1 ) \Omega_{\chi_{c1}, 2}^{(b)}]  + \epsilon_c^2 [ 2 \Omega_{\chi_{c1}, 1}^{(c)} + ( w - 1 ) \Omega_{\chi_{c1}, 2}^{(c)} ] \Big]  \notag \\
&  + \frac{1}{4 \sqrt{2}} \epsilon_b \epsilon_c \Big[ \tilde{\Lambda} [ 2 ( w - 3 ) \Sigma_{\chi_{c1}, 1}^{(b)} + ( w - 1 )^2 \Sigma_{\chi_{c1}, 2}^{(b)} ] - \tilde{\Lambda}' [ 2 ( w + 1 ) \Sigma_{\chi_{c1}, 1}^{(c)} - ( w - 1 )^2 \Sigma_{\chi_{c1}, 2}^{(c)} ]   \notag \\
& + [ 2 \Omega_{\chi_{c1}, 1}^{(Q), \text{mix}} - ( w - 1 ) \Omega_{\chi_{c1}, 2}^{(Q), \text{mix}} ] \Big] \,, 
\end{align}
\begin{align}
g_{V_3} & = \frac{w + 1}{2 \sqrt{2}} \Xi - \frac{1}{2 \sqrt{2}} \Big[ \epsilon_b [ 2 \Sigma_{\chi_{c1}, 1}^{(b)} + ( w + 1 ) \Sigma_{\chi_{c1}, 2}^{(b)} ] - \epsilon_c [ 2 \Sigma_{\chi_{c1}, 1}^{(c)} - ( w + 1 ) \Sigma_{\chi_{c1}, 2}^{(c)} ] \Big]  \notag \\
&  + \frac{1}{4 \sqrt{2}} \Big[ \epsilon_b [ 2 \Upsilon_{2, \chi_{c1}, 1} ^{(b)} - ( w + 1 ) \Upsilon_{2, \chi_{c1}, 2}^{(b)} ] - \epsilon_c ( w + 1 ) [ 2 \Upsilon_{2, \chi_{c1}, 1} ^{(c)} - \Upsilon_{2, \chi_{c1}, 2}^{(c)} ] \Big]  \notag \\
&  + \frac{1}{\sqrt{2}} \Big[ \epsilon_b^2 [ (w + 1) \Upsilon_{1A} ^{(b)} - 2 \Upsilon_{1B} ^{(b)} ] + \epsilon_c^2 [ (w + 1) \Upsilon_{1A} ^{(c)} - 2 \Upsilon_{1B} ^{(c)} ] \Big] \notag \\
& - \frac{1}{2 \sqrt{2}} \Big[ \epsilon_b^2 [ 2 \Omega_{\chi_{c1}, 1}^{(b)} - ( w + 1 ) \Omega_{\chi_{c1}, 2}^{(b)} ] + \epsilon_c^2 [ 2 \Omega_{\chi_{c1}, 1}^{(c)} + ( w + 1 ) \Omega_{\chi_{c1}, 2}^{(c)} ] \Big]  \notag \\
& - \frac{1}{4 \sqrt{2}} \epsilon_b \epsilon_c \Bigg[
 ( w + 1 ) \Big( \tilde{\Lambda} [ 2 \Sigma_{\chi_{c1}, 1}^{(b)} + ( w - 1 ) \Sigma_{\chi_{c1}, 2}^{(b)} ] - 
 \tilde{\Lambda}' [ 2 \Sigma_{\chi_{c1}, 1}^{(c)} - ( w - 1 ) \Sigma_{\chi_{c1}, 2}^{(c)}  ] \Big) 
  \notag \\
& + [ 2 \Omega_{\chi_{c1}, 1}^{(Q), \text{mix}} - ( w + 1 ) \Omega_{\chi_{c1}, 2}^{(Q), \text{mix}} ] \Bigg]
\;, \\
g_{A} & = \frac{w + 1}{\sqrt{2}} \Xi - \frac{1}{\sqrt{2}} \Big[ \epsilon_b [ 2 \Sigma_{\chi_{c1}, 1}^{(b)} + ( w - 1 ) \Sigma_{\chi_{c1}, 2}^{(b)} ] - \epsilon_c [ 2 \Sigma_{\chi_{c1}, 1}^{(c)} - ( w - 1 ) \Sigma_{\chi_{c1}, 2}^{(c)} ] \Big]  \notag \\
&  + \frac{1}{2 \sqrt{2}} \Big[ \epsilon_b [ 2 \Upsilon_{2, \chi_{c1}, 1} ^{(b)} - ( w + 1 ) \Upsilon_{2, \chi_{c1}, 2}^{(b)} ] - \epsilon_c ( w + 1 ) [ 2 \Upsilon_{2, \chi_{c1}, 1} ^{(c)} - \Upsilon_{2, \chi_{c1}, 2}^{(c)} ] \Big]  \notag \\
&  + \sqrt{2} \Big[ \epsilon_b^2 [ (w + 1) \Upsilon_{1A} ^{(b)} - 2 \Upsilon_{1B} ^{(b)} ] + \epsilon_c^2 [ (w + 1) \Upsilon_{1A} ^{(c)} - 2 \Upsilon_{1B} ^{(c)} ] \Big]
\notag \\
&  - \frac{1}{\sqrt{2}} \Big[ \epsilon_b^2 [ 2 \Omega_{\chi_{c1}, 1}^{(b)} - ( w - 1 ) \Omega_{\chi_{c1}, 2}^{(b)} ] + \epsilon_c^2 [ 2 \Omega_{\chi_{c1}, 1}^{(c)} + ( w - 1 ) \Omega_{\chi_{c1}, 2}^{(c)} ] \Big]  \notag \\
&  - \frac{1}{2 \sqrt{2}} \epsilon_b \epsilon_c \Big[ ( w + 1 ) \Big( \tilde{\Lambda} [ 2 \Sigma_{\chi_{c1}, 1}^{(b)} + ( w - 1 ) \Sigma_{\chi_{c1}, 2}^{(b)} ] - \tilde{\Lambda}' [ 2 \Sigma_{\chi_{c1}, 1}^{(c)} - ( w - 1 ) \Sigma_{\chi_{c1}, 2}^{(c)} ] \Big) \Big]  \notag \\
&  + ( 2 \Omega_{\chi_{c1}, 1}^{(Q), \text{mix}} - ( w + 1 ) \Omega_{\chi_{c1}, 2}^{(Q), \text{mix}} ) \Big]  
\;, \\
g_S & = \sqrt{2} \big[ \epsilon_b \Sigma_{\chi_{c1}, 1}^{(b)} + \epsilon_c \Sigma_{\chi_{c1}, 1}^{(c)} \big] - \frac{1}{\sqrt{2}} \big[ \epsilon_b \Upsilon_{2, \chi_{c1}, 1}^{(b)} - \epsilon_c ( w + 1 ) \Upsilon_{2, \chi_{c1}, 1}^{(c)} \big] \notag \\
&  + 2 \sqrt{2} \big[ \epsilon_b^2 \Upsilon_{1B}^{(b)} + \epsilon_c^2 \Upsilon_{1B}^{(c)} \big]+ \sqrt{2} \big[ \epsilon_b^2 \Omega_{\chi_{c1}, 1}^{(b)} - \epsilon_c^2 \Omega_{\chi_{c1}, 1}^{(c)} \big]  \notag \\
&  + \frac{1}{\sqrt{2}} \epsilon_b \epsilon_c \big[ ( w + 1 )( \tilde{\Lambda} \Sigma_{\chi_{c1}, 1}^{(b)} + \tilde{\Lambda}' \Sigma_{\chi_{c1}, 1}^{(c)} ) -\Omega_{\chi_{c1}, 1}^{(Q), \text{mix}} \big] \;, 
\end{align}
\begin{align}
g_{T_1} & = - \frac{1}{\sqrt{2}} \Big[ \epsilon_b [ 2 \Sigma_{\chi_{c1}, 1}^{(b)} + ( w - 1 ) \Sigma_{\chi_{c1}, 2}^{(b)} ] + \epsilon_c [ 2 \Sigma_{\chi_{c1}, 1}^{(c)} - ( w - 1 ) \Sigma_{\chi_{c1}, 2}^{(c)} ] \Big]  \notag \\
& - \frac{1}{\sqrt{2}} \Big[ \epsilon_b^2 [ 2 \Omega_{\chi_{c1}, 1}^{(b)} - ( w - 1 ) \Omega_{\chi_{c1}, 2}^{(b)} ] - \epsilon_c^2 [ 2 \Omega_{\chi_{c1}, 1}^{(c)} + ( w - 1 ) \Omega_{\chi_{c1}, 2}^{(c)} ] \Big]
\;, \\
g_{T_2} & = \frac{w + 1}{\sqrt{2}} \Xi  + \frac{1}{2 \sqrt{2}} \Big[ \epsilon_b [ 2 \Upsilon_{2, \chi_{c1}, 1} ^{(b)} - ( w + 1 ) \Upsilon_{2, \chi_{c1}, 2}^{(b)} ] - \epsilon_c ( w + 1 ) [ 2 \Upsilon_{2, \chi_{c1}, 1} ^{(c)} - \Upsilon_{2, \chi_{c1}, 2}^{(c)} ] \Big]  \notag \\
&  + \sqrt{2} \Big[ \epsilon_b^2 [ (w + 1) \Upsilon_{1A} ^{(b)} - 2 \Upsilon_{1B} ^{(b)} ] + \epsilon_c^2 [ (w + 1) \Upsilon_{1A} ^{(c)} - 2 \Upsilon_{1B} ^{(c)} ] \Big]
\notag \\
&
+ \frac{1}{2 \sqrt{2}} \epsilon_b \epsilon_c \Big[ ( w + 1 ) \Big( \tilde{\Lambda} [ 2 \Sigma_{\chi_{c1}, 1}^{(b)} + ( w - 1 ) \Sigma_{\chi_{c1}, 2}^{(b)} ] - \tilde{\Lambda}' [ 2 \Sigma_{\chi_{c1}, 1}^{(c)} - ( w - 1 ) \Sigma_{\chi_{c1}, 2}^{(c)} ] \Big)   \notag \\
& + [ 2 \Omega_{\chi_{c1}, 1}^{(Q), \text{mix}} - ( w + 1 ) \Omega_{\chi_{c1}, 2}^{(Q), \text{mix}} ] \Big]   
\;, \\
g_{T_3} & = \frac{1}{\sqrt{2}} \Xi - \frac{1}{\sqrt{2}} \big[ \epsilon_b \Sigma_{\chi_{c1}, 2}^{(b)} - \epsilon_c \Sigma_{\chi_{c1}, 2}^{(c)} \big]  - \frac{1}{2 \sqrt{2}} \big[ \epsilon_b \Upsilon_{2, \chi_{c1}, 2}^{(b)} - \epsilon_c \Upsilon_{2, \chi_{c1}, 2}^{(c)} \big] \notag \\
& + \sqrt{2} \big[ \epsilon_b^2 \Upsilon_{1A} ^{(b)} + \epsilon_c^2 \Upsilon_{1A} ^{(c)} \big]
+ \frac{1}{\sqrt{2}} \big[ \epsilon_b^2 \Omega_{\chi_{c1}, 2}^{(b)} + \epsilon_c^2 \Omega_{\chi_{c1}, 2}^{(c)} \big]  \notag \\
&  + \frac{1}{2 \sqrt{2}} \epsilon_b \epsilon_c \Big[ \tilde{\Lambda} [ 4 \Sigma_{\chi_{c1}, 1}^{(b)} + ( w - 1 ) \Sigma_{\chi_{c1}, 2}^{(b)} ] +\tilde{\Lambda}' ( w - 1 ) \Sigma_{\chi_{c1}, 2}^{(c)}   - \Omega_{\chi_{c1}, 2}^{(Q), \text{mix}} \Big] 
\;.
\end{align}

\subsection{$B_c \to \chi_{c2}$}

We define: 
\begin{align}
\Sigma_{\chi_{c2}}^{(b)} & = \Sigma_1^{(b)} + ( w + 1 ) \Sigma_3^{(b)} - 3 \Sigma_4^{(b)} - \Sigma_7^{(b)}
\;, \\
\Sigma_{\chi_{c2}, 1}^{(c)} & = ( w + 1 ) \Sigma_2^{(c)} - \Sigma_4^{(c)}
\;, \\
\Sigma_{\chi_{c2}, 2}^{(c)} & = ( w + 1 ) \Sigma_2^{(c)} + \Sigma_4^{(c)}
\;, \\
\Omega_{\chi_{c2}}^{(b)} & = \Omega_3^{(b)} + ( w + 1 ) [ \Omega_{4}^{(b)} + \Omega_{10}^{(b)} + w \Omega_{12}^{(b)} - \Omega_{18}^{(b)} - \Omega_{24}^{(b)} ] + w \Omega_5^{(b)}  \notag \\
& - 3 [ \Omega_{6}^{(b)} + \Omega_{13}^{(b)} + w \Omega_{16}^{(b)} + \Omega_{22}^{(b)} ] - \Omega_8^{(b)} + \Omega_{23}^{(b)} + w \Omega_{25}^{(b)}
\;, \\
\Omega_{\chi_{c2}, 1}^{(c)} & = - (w + 1) [ w \Omega_{9}^{(c)} + \Omega_{10}^{(c)} - \Omega_{13}^{(c)} ] + w \Omega_{14}^{(c)} + \Omega_{18}^{(c)} + \Omega_{22}^{(c)}
\;, \\
\Omega_{\chi_{c2}, 2}^{(c)} & = (w + 1) [ w \Omega_{9}^{(c)} + \Omega_{10}^{(c)} - \Omega_{13}^{(c)} ] + w \Omega_{14}^{(c)} + \Omega_{18}^{(c)} + \Omega_{22}^{(c)}
\;, \\
 \Omega_{\chi_{c2}, 1}^{(Q), \text{mix}} & = \Omega_{8}^{(Q)} - w \Omega_{2}^{(Q)} + ( 1 - w ) [ \Omega_{3}^{(Q)} + (1 + w) \Omega_{10}^{(Q)}  \notag \\
&   - 3 \Omega_{13}^{(Q)} - \Omega_{18}^{(Q)} + \Omega_{23}^{(Q)} ] + (w - 3) \Omega_{22}^{(Q)}
\;, \\
 \Omega_{\chi_{c2}, 2}^{(Q), \text{mix}}& =  - \Omega_{8}^{(Q)} + (2 - w) \Omega_{2}^{(Q)} + ( 1 - w ) [ \Omega_{3}^{(Q)} 
 + (1 + w) \Omega_{10}^{(Q)}  \notag \\
&   - 3 \Omega_{13}^{(Q)} + \Omega_{18}^{(Q)} + \Omega_{23}^{(Q)}] +(1+w) \Omega_{22}^{(Q)}
\;, \\
\Upsilon_{2, \chi_{c2}}^{(b)} & = \Upsilon_{2B}^{(b)} - 2 \Upsilon_{2C}^{(b)} - 2 (w-1 ) \Upsilon_{2F}^{(b)} - \Upsilon_{2I}^{(b)} - 3 \Upsilon_{2J}^{(b)}
\;, \\
\Upsilon_{2, \chi_{c2}, 1}^{(c)} & = \Upsilon_{2A}^{(c)} - \Upsilon_{2H}^{(c)} + \Upsilon_{2J}^{(c)}
\;, \\
\Upsilon_{2, \chi_{c2}, 2}^{(c)}& = \Upsilon_{2A}^{(c)}  - 2 (w-1) \Upsilon_{2E}^{(c)}  - \Upsilon_{2H}^{(c)} + \Upsilon_{2J}^{(c)}
\;.
\end{align}
%
With these definitions the form factors are expressed in terms of universal functions:
\begin{align}
k_V & = - \Xi+ \big[ \epsilon_b \Sigma_{\chi_{c_2}}^{(b)} + \epsilon_c \Sigma_{\chi_{c_2}, 1}^{(c)} \big] + \frac{1}{2} \big[ \epsilon_b \Upsilon_{2, \chi_{c_2}}^{(b)} + \epsilon_c \Upsilon_{2, \chi_{c_2}, 1}^{(c)} \big] - 2 \big[ \epsilon_b^2 \Upsilon_{1A}^{(b)} + \epsilon_c^2 \Upsilon_{1A}^{(c)} \big]
\notag \\ 
&  - \big[ \epsilon_b^2 \Omega_{\chi_{c_2}}^{(b)} - \epsilon_c^2 \Omega_{\chi_{c_2}, 1}^{(c)} \big] + \frac{1}{2} \epsilon_b \epsilon_c \Big[ ( w - 1 ) \Big( \tilde{\Lambda} \Sigma_{\chi_{c_2}}^{(b)} + \tilde{\Lambda}' \Sigma_{\chi_{c_2}, 1}^{(c)} \Big) + \Omega_{\chi_{c_2}, 1}^{(Q), \text{mix}} \Big] 
\;, \\
k_{A_1} & = ( w + 1 ) \Xi - ( w - 1 ) \big[ \epsilon_b \Sigma_{\chi_{c_2}}^{(b)} + \epsilon_c \Sigma_{\chi_{c_2}, 1}^{(c)} \big] \notag \\
&  - \frac{w + 1}{2} \big[ \epsilon_b \Upsilon_{2, \chi_{c_2}}^{(b)} + \epsilon_c \Upsilon_{2, \chi_{c_2}, 1}^{(c)} \big] + 2 ( w + 1 ) \big[ \epsilon_b^2 \Upsilon_{1A}^{(b)} + \epsilon_c^2 \Upsilon_{1A}^{(c)} \big]
\notag \\ 
& + ( w - 1 ) \big[ \epsilon_b^2 \Omega_{\chi_{c_2}}^{(b)} - \epsilon_c^2 \Omega_{\chi_{c_2}, 1}^{(c)} \big]  - \frac{w + 1}{2} \epsilon_b \epsilon_c \big[ ( w - 1 ) \big( \tilde{\Lambda} \Sigma_{\chi_{c_2}}^{(b)} + \tilde{\Lambda}' \Sigma_{\chi_{c_2},  1}^{(c)} \big) + \Omega_{\chi_{c_2}, 1}^{(Q), \text{mix}} \big] 
\;, \\
k_{A_2} & = - \frac{\epsilon_c}{w + 1} \big[ \Sigma_{\chi_{c_2}, 1}^{(c)} + \Sigma_{\chi_{c_2}, 2}^{(c)} \big] - \frac{\epsilon_c}{2 ( w - 1 )} \big[ \Upsilon_{2, \chi_{c_2}, 1}^{(c)} - \Upsilon_{2, \chi_{c_2}, 2}^{(c)} \big]
- \frac{\epsilon_c^2}{w + 1} \big[ \Omega_{\chi_{c_2}, 1}^{(c)} - \Omega_{\chi_{c_2}, 2}^{(c)} \big]  \notag \\
&  - \frac{\epsilon_b \epsilon_c}{2 ( w - 1 )} \Big[ ( w - 1 ) \Big( 2 \tilde{\Lambda} \Sigma_{\chi_{c_2}}^{(b)} + \tilde{\Lambda}' ( \Sigma_{\chi_{c_2}, 1}^{(c)} + \Sigma_{\chi_{c_2}, 2}^{(c)} ) \Big) + \Omega_{\chi_{c_2}, 1}^{(Q), \text{mix}} + \Omega_{\chi_{c_2}, 2}^{(Q), \text{mix}} \Big] 
\;, \\
k_{A_3} & = - \Xi + \Big[ \epsilon_b \Sigma_{\chi_{c_2}}^{(b)} + \frac{\epsilon_c}{w + 1} \Big( w \Sigma_{\chi_{c_2}, 1}^{(c)} - \Sigma_{\chi_{c_2}, 2}^{(c)}\Big) \Big] + \frac{1}{2} \Big[ \epsilon_b \Upsilon_{2, \chi_{c_2}}^{(b)} + \frac{\epsilon_c}{w - 1} \Big( w \Upsilon_{2, \chi_{c_2}, 1}^{(c)} - \Upsilon_{2, \chi_{c_2}, 2}^{(c)} \Big) \Big] \notag \\ 
&  - 2 \big[ \epsilon_b^2 \Upsilon_{1A}^{(b)} + \epsilon_c^2 \Upsilon_{1A}^{(c)} \big]
 - \Big[ \epsilon_b^2 \Omega_{\chi_{c_2}}^{(b)}- \frac{\epsilon_c^2}{w + 1} \Big( w \Omega_{\chi_{c_2}, 1}^{(c)} + \Omega_{\chi_{c_2}, 2}^{(c)} \Big) \Big]  \notag \\
&  + \frac{\epsilon_b \epsilon_c}{2 ( w - 1 )} \Big[ ( w - 1 ) \Big( ( w + 1 ) \tilde{\Lambda} \Sigma_{\chi_{c_2}}^{(b)} + \tilde{\Lambda}' ( w \Sigma_{\chi_{c_2}, 1}^{(c)} + \Sigma_{\chi_{c_2}, 2}^{(c)} ) \Big) + w \Omega_{\chi_{c_2}, 1}^{(Q), \text{mix}} + \Omega_{\chi_{c_2}, 2}^{(Q), \text{mix}} \Big]
\;, \\
k_P & = - \Xi + \big[ \epsilon_b \Sigma_{\chi_{c_2}}^{(b)} + \epsilon_c \Sigma_{\chi_{c_2}, 2}^{(c)} \big] + \frac{1}{2} \big[ \epsilon_b \Upsilon_{2, \chi_{c_2}}^{(b)} + \epsilon_c \Upsilon_{2, \chi_{c_2}, 2}^{(c)} \big] - 2 \big[ \epsilon_b^2 \Upsilon_{1A}^{(b)} + \epsilon_c^2 \Upsilon_{1A}^{(c)} \big]
\notag \\ 
&- \big[ \epsilon_b^2 \Omega_{\chi_{c_2}}^{(b)} + \epsilon_c^2 \Omega_{\chi_{c_2}, 2}^{(c)} \big] + \frac{1}{2} \epsilon_b \epsilon_c \big[ ( w - 1 ) \big( \tilde{\Lambda} \Sigma_{\chi_{c_2}}^{(b)} + \tilde{\Lambda}' \Sigma_{\chi_{c_2}, 2}^{(c)} \big) + \Omega_{\chi_{c_2}, 2}^{(Q), \text{mix}} \big] 
\;, \\
k_{T_1} & = - \Xi + [ \epsilon_b \Sigma_{\chi_{c_2}}^{(b)} - \epsilon_c \Sigma_{\chi_{c_2}, 1}^{(c)} ] + \frac{1}{2} [ \epsilon_b \Upsilon_{2, \chi_{c_2}}^{(b)} + \epsilon_c \Upsilon_{2, \chi_{c_2}, 1}^{(c)} ] - 2 [ \epsilon_b^2 \Upsilon_{1A}^{(b)} + \epsilon_c^2 \Upsilon_{1A}^{(c)} ]
\notag \\
&- [ \epsilon_b^2 \Omega_{\chi_{c_2}}^{(b)} + \epsilon_c^2 \Omega_{\chi_{c_2}, 1}^{(c)} ]  - \frac{1}{2} \epsilon_b \epsilon_c \Big[ ( w - 1 ) \Big( \tilde{\Lambda} \Sigma_{\chi_{c_2}}^{(b)} + \tilde{\Lambda}' \Sigma_{\chi_{c_2}, 1}^{(c)} \Big) + \Omega_{\chi_{c_2}, 1}^{(Q), \text{mix}} \Big]
\;, 
\end{align}
\begin{align}
k_{T_2} & = - \Xi - [ \epsilon_b \Sigma_{\chi_{c_2}}^{(b)} - \epsilon_c \Sigma_{\chi_{c_2}, 1}^{(c)} ]+ \frac{1}{2} [ \epsilon_b \Upsilon_{2, \chi_{c_2}}^{(b)} + \epsilon_c \Upsilon_{2, \chi_{c_2}, 1}^{(c)} ] - 2 [ \epsilon_b^2 \Upsilon_{1A}^{(b)} + \epsilon_c^2 \Upsilon_{1A}^{(c)} ] \notag \\ & + [ \epsilon_b^2 \Omega_{\chi_{c_2}}^{(b)} + \epsilon_c^2 \Omega_{\chi_{c_2}, 1}^{(c)} ] - \frac{1}{2} \epsilon_b \epsilon_c \Big[ ( w - 1 ) \Big( \tilde{\Lambda} \Sigma_{\chi_{c_2}}^{(b)} + \tilde{\Lambda}' \Sigma_{\chi_{c_2}, 1}^{(c)} \Big) + \Omega_{\chi_{c_2}, 1}^{(Q), \text{mix}} \Big] 
\;, \\
k_{T_3} & = - \frac{\epsilon_c}{w + 1} [ \Sigma_{\chi_{c_2}, 1}^{(c)} + \Sigma_{\chi_{c_2}, 2}^{(c)} ] + \frac{\epsilon_c}{2 ( w - 1 )} [ \Upsilon_{2, \chi_{c_2}, 1}^{(c)} - \Upsilon_{2, \chi_{c_2}, 2}^{(c)} ]
 - \frac{\epsilon_c^2}{w + 1} [ \Omega_{\chi_{c_2}, 1}^{(c)} - \Omega_{\chi_{c_2}, 2}^{(c)} ] \notag \\ & - \frac{\epsilon_b \epsilon_c}{2 ( w - 1 )} \Big[ ( w - 1 ) \Big( 2 \tilde{\Lambda} \Sigma_{\chi_{c_2}}^{(b)} + \tilde{\Lambda}' ( \Sigma_{\chi_{c_2}, 1}^{(c)} + \Sigma_{\chi_{c_2}, 2}^{(c)} ) \Big) + \Omega_{\chi_{c_2}, 1}^{(Q), \text{mix}} + \Omega_{\chi_{c_2}, 2}^{(Q), \text{mix}} \Big] 
\;.
\end{align}

\subsection{$B_c \to h_{c}$}

We define:
\begin{align}
\Sigma_{h_c, 1}^{(b)} & = w \Sigma_1^{(b)} + ( w^2 - 1 ) \Sigma_3^{(b)} - ( w - 1 ) [ 3 \Sigma_4^{(b)} + \Sigma_6^{(b)} ] - ( w - 3 ) \Sigma_7^{(b)}
\;, \\
\Sigma_{h_c, 2}^{(b)} & = \Sigma_1^{(b)} + ( w + 1 ) \Sigma_3^{(b)} - 3 \Sigma_4^{(b)} - \Sigma_7^{(b)}
\;, \\
\Sigma_{h_c, 1}^{(c)} & = \Sigma_1^{(c)} - ( w^2 - 1 ) \Sigma_2^{(c)} + ( w - 1 ) [ 3 \Sigma_4^{(c)} + \Sigma_5^{(c)} ] - 2 \Sigma_7^{(c)}
\;, \\
\Sigma_{h_c, 2}^{(c)} & = ( w + 1 ) \Sigma_2^{(c)} - 3 \Sigma_4^{(c)} - 2 \Sigma_5^{(c)}
\;, \\
\Omega_{h_c, 1}^{(b)} & = w [ \Omega_3^{(b)} + w \Omega_5^{(b)} - \Omega_8^{(b)} ] + ( w^2 - 1 ) [ \Omega_{4}^{(b)}  + \Omega_{10}^{(b)} + w \Omega_{12}^{(b)} - \Omega_{18}^{(b)} ]  \notag \\
&  - ( w - 1 ) [ 3 \Omega_{6}^{(b)} + 3 \Omega_{13}^{(b)} +  3 w \Omega_{16}^{(b)}  + \Omega_{17}^{(b)} + w \Omega_{20}^{(b)} + 3 \Omega_{22}^{(b)} + w \Omega_{24}^{(b)} ]  \notag \\
& + ( w - 3 ) \, [ \Omega_{23}^{(b)} + w \Omega_{25}^{(b)} ]
\;, \\
\Omega_{h_c, 2}^{(b)} & = \Omega_3^{(b)} + ( w + 1 ) [ \Omega_{4}^{(b)} + \Omega_{10}^{(b)} + w \Omega_{12}^{(b)} - \Omega_{18}^{(b)} - \Omega_{24}^{(b)} ] + w \Omega_5^{(b)} - 3 \Omega_{6}^{(b)} - \Omega_8^{(b)} - 3 \Omega_{13}^{(b)}  \notag \\
& - 3 w \Omega_{16}^{(b)} - 3 \Omega_{22}^{(b)} + \Omega_{23}^{(b)} + w \Omega_{25}^{(b)}
\;, \\
\Omega_{h_c, 1}^{(c)} & = w \Omega_{1}^{(c)} + \Omega_4^{(c)} - \Omega_6^{(c)} - ( w^2 - 1 ) [ w \Omega_{9}^{(c)} + \Omega_{10}^{(c)} - \Omega_{13}^{(c)} ]  \notag \\
&  + ( w - 1 ) [ 3 w \Omega_{14}^{(c)} + w \Omega_{15}^{(c)} + \Omega_{17}^{(c)} + 3 \Omega_{18}^{(c)} + 3 \Omega_{22}^{(c)} - \Omega_{23}^{(c)} ] - 2 w \Omega_{21}^{(c)} - 2 \Omega_{24}^{(c)}
\;, \\
\Omega_{h_c, 2}^{(c)} & = ( w + 1 ) [ w \Omega_{9}^{(c)} + \Omega_{10}^{(c)} - \Omega_{13}^{(c)} ] - 3 w \Omega_{14}^{(c)} - 2 w \Omega_{15}^{(c)}  - 2 \Omega_{17}^{(c)} - 3 \Omega_{18}^{(c)} - 3 \Omega_{22}^{(c)} + 2 \Omega_{23}^{(c)} 
\;, \\
\Omega_{h_c, 1}^{(Q), \text{mix}} & = ( w + 1 ) ( w + 2 ) \Omega_{2}^{(Q)} + ( w + 1 ) [ w \Omega_{3}^{(Q)} - \Omega_{4}^{(Q)} + 2 \Omega_{24}^{(Q)} ] + 3 \Omega_{6}^{(Q)} - ( w + 2 ) \Omega_{7}^{(Q)}  \notag \\
&  - 3 w \Omega_{8}^{(Q)} + ( w - 1 ) ( w + 1 )^2 \Omega_{10}^{(Q)} - ( w^2 - 1 ) [ 3 \Omega_{13}^{(Q)} + \Omega_{17}^{Q} + 3 \Omega_{18}^{(Q)} ]  \notag \\
&   - ( w - 7 ) ( w + 1 ) \Omega_{22}^{(Q)} + ( w - 3 ) ( w + 1 ) \Omega_{23}^{(Q)}
\;, \\
\Omega_{h_c, 2}^{(Q), \text{mix}} & = ( w + 2 ) \Omega_{2}^{(Q)} + ( w + 1 ) \Omega_{3}^{(Q)} - 2 \Omega_{7}^{(Q)} - 3 \Omega_{8}^{(Q)} + ( w^2 - 1 ) \Omega_{10}^{(Q)}  \notag \\
&  - ( w - 1 ) [ 3 \Omega_{13}^{(Q)} + 2 \Omega_{17}^{(Q)} + 3 \Omega_{18}^{(Q)} ] - ( w - 7 ) \Omega_{22}^{(Q)} + ( w - 5 ) \Omega_{23}^{(Q)}
\;, \\
\Upsilon_{2, h_c, 1}^{(b)} & = 2 w \Upsilon_{2C}^{(b)} + ( w+1) [ 2 (w - 1) \Upsilon_{2F}^{(b)} + 2  \Upsilon_{2I}^{(b)} + 3 \Upsilon_{2J}^{(b)} ]
\;, \\
\Upsilon_{2, h_c, 2}^{(b)} & = \Upsilon_{2B}^{(b)} - 2 \Upsilon_{2C}^{(b)} - 2 ( w - 1 )  \Upsilon_{2F}^{(b)} - \Upsilon_{2I}^{(b)} - 3  \Upsilon_{2J}^{(b)}
\;, \\
\Upsilon_{2, h_c, 1}^{(c)} & = 2 \Upsilon_{2C}^{(c)}  - ( w + 1 ) [ 2 (w-1)  \Upsilon_{2E}^{(c)}  + 2  \Upsilon_{2H}^{(c)} -3 \Upsilon_{2J}^{(c)} ]
\;, \\
\Upsilon_{2, h_c, 2}^{(c)} & = \Upsilon_{2A}^{(c)}  - 2 (w-1)  \Upsilon_{2E}^{(c)}  - 3  \Upsilon_{2H}^{(c)} + 3 \Upsilon_{2J}^{(c)}
\;.
\end{align}
%
With these definitions we obtain:
\begin{align}
f_{V_1} & = - ( w + 1 ) \Big[ \epsilon_b \Big( \Sigma_{h_c, 1}^{(b)} - ( w - 1 ) \Sigma_{h_c, 2}^{(b)} \Big) + \epsilon_c \Big( \Sigma_{h_c, 1}^{(c)} + ( w - 1 ) \Sigma_{h_c, 2}^{(c)} \Big) \Big]  \notag \\
& + \frac{w - 1}{2} \Big[ \epsilon_b \Big( \Upsilon_{2, h_c, 1}^{(b)} + ( w + 1 ) \Upsilon_{2, h_c, 2}^{(b)} \Big) + \epsilon_c \Big( \Upsilon_{2, h_c, 1}^{(c)} - ( w + 1 ) \Upsilon_{2, h_c, 2}^{(c)} \Big) \Big]  \notag \\
&  - 2 ( w - 1 ) \big[ \epsilon_b^2 \Upsilon_{1B}^{(b)} + \epsilon_c^2 \Upsilon_{1B}^{(c)} \big] \notag \\
&  + ( w + 1 ) \Big[ \epsilon_b^2 \Big( \Omega_{h_c, 1}^{(b)} - ( w - 1 ) \Omega_{h_c, 2}^{(b)} \Big) + \epsilon_c^2 \Big( \Omega_{h_c, 1}^{(c)} + ( w - 1 ) \Omega_{h_c, 2}^{(c)} \Big) \Big]  \notag \\
&  - \frac{w - 1}{2} \epsilon_b \epsilon_c \Big[ ( w + 1 ) \Big( \tilde{\Lambda} ( \Sigma_{h_c, 1}^{(b)} - ( w - 1 ) \Sigma_{h_c, 2}^{(b)} ) + \tilde{\Lambda}' ( \Sigma_{h_c, 1}^{(c)} + ( w - 1 ) \Sigma_{h_c, 2}^{(c)} ) \Big)  \notag \\
&  + \Omega_{h_c, 1}^{(Q), \text{mix}} - ( w + 1 ) \Omega_{h_c, 2}^{(Q), \text{mix}} \Big]\;, \\
f_{V_2} & = - \Xi + \frac{1}{2} \Big[ \epsilon_b \Big( \Sigma_{h_c, 1}^{(b)} - ( w - 1 ) \Sigma_{h_c, 2}^{(b)} \Big) + \epsilon_c \Big( \Sigma_{h_c, 1}^{(c)} + ( w - 1 ) \Sigma_{h_c, 2}^{(c)} \Big) \Big]  \notag \\
&  - \frac{1}{4} \Big[ \epsilon_b \Big( \Upsilon_{2, h_c, 1}^{(b)} + ( w - 1 ) \Upsilon_{2, h_c, 2}^{(b)} \Big) + \epsilon_c \Big( \Upsilon_{2, h_c, 1}^{(c)} - ( w - 1 ) \Upsilon_{2, h_c, 2}^{(c)} \Big) \Big]  \notag \\
&  - \Big[ \epsilon_b^2 \Big( 2 \Upsilon_{1A}^{(b)} - \Upsilon_{B}^{(b)} \Big) + \epsilon_c^2 \Big( 2 \Upsilon_{1A}^{(c)} - \Upsilon_{1B}^{(c)} \Big) \Big] \notag \\
& - \frac{1}{2} \Big[ \epsilon_b^2 \Big( \Omega_{h_c, 1}^{(b)} - ( w - 1 ) \Omega_{h_c, 2}^{(b)} \Big) + \epsilon_c^2 \Big( \Omega_{h_c, 1}^{(c)} + ( w - 1 ) \Omega_{h_c, 2}^{(c)} \Big) \Big]  \notag \\
&  + \frac{1}{4} \epsilon_b \epsilon_c \Big[ \tilde{\Lambda} \Big( ( w - 3 ) \Sigma_{h_c, 1}^{(b)} - ( w - 1 )^2 \Sigma_{h_c, 2}^{(b)} \Big) + \tilde{\Lambda}' \Big( (w + 1) \Sigma_{h_c, 1}^{(c)} + ( w - 1 )^2 \Sigma_{h_c, 2}^{(c)}  \Big)  \notag \\
& + \Omega_{h_c, 1}^{(Q), \text{mix}} - ( w - 1 ) \Omega_{h_c, 2}^{(Q), \text{mix}} \Big] 
\;, \\
f_{V_3} & = - \frac{1}{2} \Big[ \epsilon_b \Big( \Sigma_{h_c, 1}^{(b)} - ( w + 1 ) \Sigma_{h_c, 2}^{(b)} \Big) + \epsilon_c \Big( \Sigma_{h_c, 1}^{(c)} + ( w + 1 ) \Sigma_{h_c, 2}^{(c)} \Big) \Big]  \notag \\
& + \frac{1}{4} \Big[ \epsilon_b \Big( \Upsilon_{2, h_c, 1}^{(b)} + ( w + 1 ) \Upsilon_{2, h_c, 2}^{(b)} \Big) + \epsilon_c \Big( \Upsilon_{2, h_c, 1}^{(c)} - ( w + 1 ) \Upsilon_{2, h_c, 2}^{(c)} \Big) \Big]  \notag \\
&  - \big[ \epsilon_b^2 \Upsilon_{1B}^{(b)} + \epsilon_c^2 \Upsilon_{1B}^{(c)} \big] \notag \\
&  + \frac{1}{2} \Big[ \epsilon_b^2 \Big( \Omega_{h_c, 1}^{(b)} - ( w + 1 ) \Omega_{h_c, 2}^{(b)} \Big) + \epsilon_c^2 \Big( \Omega_{h_c, 1}^{(c)} + ( w + 1 ) \Omega_{h_c, 2}^{(c)} \Big) \Big]  \notag \\
&  - \frac{1}{4} \epsilon_b \epsilon_c \Big[ ( w + 1 ) \Big( \tilde{\Lambda} ( \Sigma_{h_c, 1}^{(b)} - ( w - 1 ) \Sigma_{h_c, 2}^{(b)} ) + \tilde{\Lambda}' ( \Sigma_{h_c, 1}^{(c)} + ( w - 1 ) \Sigma_{h_c, 2}^{(c)} ) \Big)  \notag \\
&  + \Omega_{h_c, 1}^{(Q), \text{mix}} - ( w + 1 ) \Omega_{h_c, 2}^{(Q), \text{mix}} \Big] 
\;, 
\end{align}
\begin{align}
f_A & = \Big[ \epsilon_b \Big( \Sigma_{h_c, 1}^{(b)} - ( w - 1 ) \Sigma_{h_c, 2}^{(b)} \Big) + \epsilon_c \Big( \Sigma_{h_c, 1}^{(c)} + ( w - 1 ) \Sigma_{h_c, 2}^{(c)} \Big) \Big]  \notag \\
&  - \frac{1}{2} \Big[ \epsilon_b \Big( \Upsilon_{2, h_c, 1}^{(b)} + ( w + 1 ) \Upsilon_{2, h_c, 2}^{(b)} \Big) + \epsilon_c \Big( \Upsilon_{2, h_c, 1}^{(c)} - ( w + 1 ) \Upsilon_{2, h_c, 2}^{(c)} \Big) \Big] + 2 \big[ \epsilon_b^2 \Upsilon_{1B}^{(b)} + \epsilon_c^2 \Upsilon_{1B}^{(c)} \big]
\notag \\
&  - \Big[ \epsilon_b^2 \Big( \Omega_{h_c, 1}^{(b)} - ( w - 1 ) \Omega_{h_c, 2}^{(b)} \Big) + \epsilon_c^2 \Big( \Omega_{h_c, 1}^{(c)} + ( w - 1 ) \Omega_{h_c, 2}^{(c)} \Big) \Big]  \notag \\
&  + \frac{1}{2} \epsilon_b \epsilon_c \Big[ ( w + 1 ) \Big( \tilde{\Lambda} ( \Sigma_{h_c, 1}^{(b)} - ( w - 1 ) \Sigma_{h_c, 2}^{(b)} ) + \tilde{\Lambda}' ( \Sigma_{h_c, 1}^{(c)} + ( w - 1 ) \Sigma_{h_c, 2}^{(c)} ) \Big)  \notag \\
&  + \Omega_{h_c, 1}^{(Q), \text{mix}} - ( w + 1 ) \Omega_{h_c, 2}^{(Q), \text{mix}} \Big] 
\;, \\
f_S & = - ( w + 1 ) \Xi + \big[ \epsilon_b \Sigma_{h_c, 1}^{(b)} - \epsilon_c \Sigma_{h_c, 1}^{(c)} \big]- \frac{1}{2} \big[ \epsilon_b \Upsilon_{2, h_c, 1}^{(b)} + \epsilon_c \Upsilon_{2, h_c, 1}^{(c)} \big] 
\notag \\ 
&  - 2 \Big[ \epsilon_b^2 \Big ( w + 1 ) \Upsilon_{1A}^{(b)} - \Upsilon_{1B}^{(b)} \Big) + \epsilon_c^2 \Big( ( w + 1 ) \Upsilon_{1A}^{(c)} - \Upsilon_{1B}^{(c)} \Big) \Big]
 - \big[ \epsilon_b^2 \Omega_{h_c, 1}^{(b)} - \epsilon_c^2 \Omega_{h_c, 1}^{(c)} \big]  \notag \\
&  + \frac{1}{2} \epsilon_b \epsilon_c \Big[ ( w + 1 ) \Big( \tilde{\Lambda} \Sigma_{h_c, 1}^{(b)} - \tilde{\Lambda}' \Sigma_{h_c, 1}^{(c)} \Big) - \Omega_{h_c, 1}^{(Q), \text{mix}} \Big] 
\;, \\
f_{T_1} & = \Big[ \epsilon_b \Big( \Sigma_{h_c, 1}^{(b)} - ( w - 1 ) \Sigma_{h_c, 2}^{(b)} \Big) - \epsilon_c \Big( \Sigma_{h_c, 1}^{(c)} + ( w - 1 ) \Sigma_{h_c, 2}^{(c)} \Big) \Big]  \notag \\
&  - \Big[ \epsilon_b^2 \Big( \Omega_{h_c, 1}^{(b)} - ( w - 1 ) \Omega_{h_c, 2}^{(b)} \Big) - \epsilon_c^2 \Big( \Omega_{h_c, 1}^{(c)} + ( w - 1 ) \Omega_{h_c, 2}^{(c)} \Big) \Big]
\;, \\
f_{T_2} & =  - \frac{1}{2} \Big[ \epsilon_b \Big( \Upsilon_{2, h_c, 1}^{(b)} + ( w + 1 ) \Upsilon_{2, h_c, 2}^{(b)} \Big) + \epsilon_c \Big( \Upsilon_{2, h_c, 1}^{(c)} - ( w + 1 ) \Upsilon_{2, h_c, 2}^{(c)} \Big) \Big] + 2 \big[ \epsilon_b^2 \Upsilon_{1B}^{(b)} + \epsilon_c^2 \Upsilon_{1B}^{(c)} \big]
\notag \\ &\,\,\,\,
 - \frac{1}{2} \epsilon_b \epsilon_c \Big[ ( w + 1 ) \Big( \tilde{\Lambda} ( \Sigma_{h_c, 1}^{(b)} - ( w - 1 ) \Sigma_{h_c, 2}^{(b)} ) + \tilde{\Lambda}' ( \Sigma_{h_c, 1}^{(c)} + ( w - 1 ) \Sigma_{h_c, 2}^{(c)} ) \Big)  \notag \\
&  + \Omega_{h_c, 1}^{(Q), \text{mix}} - ( w + 1 ) \Omega_{h_c, 2}^{(Q), \text{mix}} \Big] 
\;, \\
f_{T_3} & = \Xi - \big[ \epsilon_b \Sigma_{h_c, 2}^{(b)} + \epsilon_c \Sigma_{h_c, 2}^{(c)} \big] 
 - \frac{1}{2} \big[ \epsilon_b \Upsilon_{2, h_c, 2}^{(b)} - \epsilon_c\Upsilon_{2, h_c, 2}^{(c)} \big] + 2 [ \epsilon_b^2 \Upsilon_{1A}^{(b)} + \epsilon_c^2 \Upsilon_{1A}^{(c)} ]\notag \\ &
  + \big[ \epsilon_b^2 \Omega_{h_c, 2}^{(b)} + \epsilon_c^2 \Omega_{h_c, 2}^{(c)} \big]  
  - \frac{1}{2} \epsilon_b \epsilon_c \Big[ \tilde{\Lambda} \Big( 2 \Sigma_{h_c, 1}^{(b)} - ( w - 1 ) \Sigma_{h_c, 2}^{(b)} \Big) + ( w - 1 ) \tilde{\Lambda}' \Sigma_{h_c, 2}^{(c)} - \Omega_{h_c, 2}^{(Q), \text{mix}} \Big] 
\;.
\end{align}

\section{A set of relations among universal functions}\label{app:relations}
\numberwithin{equation}{section}
In this Appendix we use the definitions in Eqs.~(\ref{Sigmas}), (\ref{Omegas}) and (\ref{upsilon2}), together with the constraints \eqref{relations_among_Sigmas}, \eqref{relations_among_psi},  to relate the various structures entering in the  expressions  (B.1.1)-(B.1.7), (B.2.1)-(B.2.14), (B.3.1)-(B.3.11) and (B.4.1)-(B.4.14). The results allow us to obtain the relations among  form factors  in Sec.~\ref{sec4}. 

\bea
\Sigma_{\chi_{c_0}}^{(b)} & =& 3 \Sigma_{\chi_{c_1}, 1}^{(b)} + ( w - 1 ) \Sigma_{\chi_{c_1}, 2}^{(b)}  \nn \\
\Sigma_{\chi_{c2}}^{(b)}&=&\Sigma_{\chi_{c1},2}^{(b)} \nn \\
\Sigma_{h_c,1}^{(b)}&=&\Sigma_{\chi_{c1},1}^{(b)} + ( w - 1 ) \Sigma_{\chi_{c_1}, 2}^{(b)}   \\
\Sigma_{h_c,2}^{(b)}&=&\Sigma_{\chi_{c1},2}^{(b)} \nn
\eea

\bea
\Sigma_{h_c,1}^{(c)} &=& - \frac{1}{3} \Sigma_{\chi_{c0}}^{(c)} + 2 \Sigma_{\chi_{c1},1}^{(c)} - ( w - 1 ) \Sigma_{\chi_{c1},2}^{(c)} - \frac{w - 1}{3} \big[ 3 \Sigma_{\chi_{c2},1}^{(c)} - 2 \Sigma_{\chi_{c2},2}^{(c)} \big]
\nn \\
\Sigma_{h_c,2}^{(c)} &=& \frac{1}{2} \Sigma_{\chi_{c1},2}^{(c)} + \frac{1}{2} \big[ 3 \Sigma_{\chi_{c2},1}^{(c)} - 2 \Sigma_{\chi_{c2},2}^{(c)} \big]
\eea

\bea
\Upsilon_{2,\chi_{c0}}^{(b)} &=& 3 \Upsilon_{2,\chi_{c1},1}^{(b)} - (w + 1) \Upsilon_{2,\chi_{c1},2}^{(b)}
\nn \\
\Upsilon_{2,\chi_{c2}}^{(b)} &=& \Upsilon_{2,\chi_{c1},2}^{(b)}
\nn \\
\Upsilon_{2,h_c,1}^{(b)} &=& \Upsilon_{2,\chi_{c1},1}^{(b)} - (w + 1) \Upsilon_{2,\chi_{c1},2}^{(b)}
 \\
\Upsilon_{2,h_c,2}^{(b)} &=& \Upsilon_{2,\chi_{c1},2}^{(b)} \nn
\eea

\bea
\Upsilon_{2,h_c,1}^{(c)} &=& - \frac{1}{3} \Upsilon_{2,\chi_{c0}}^{(c)} + (w + 1) \big[ 2 \Upsilon_{2,\chi_{c1},1}^{(c)} - \Upsilon_{2,\chi_{c1},2}^{(c)} \big] + \frac{w + 1}{3} \big[ 3 \Upsilon_{2,\chi_{c2},1}^{(c)} + 2 \Upsilon_{2,\chi_{c2},2}^{(c)} \big]
\nn \\
\Upsilon_{2,h_c,2}^{(c)} &=& - \frac{1}{2} \Upsilon_{2,\chi_{c1},2}^{(c)} + \frac{1}{2} \big[ 3 \Upsilon_{2,\chi_{c2},1}^{(c)} + 2 \Upsilon_{2,\chi_{c2},2}^{(c)} \big]
\eea

\bea
\Omega_{\chi_{c_0}}^{(b)} &=& - 3 \Omega_{\chi_{c_1}, 1}^{(b)} + ( w - 1 ) \Omega_{\chi_{c_1}, 2}^{(b)}
\nn \\
\Omega_{\chi_{c2}}^{(b)} &=& \Omega_{\chi_{c1},2}^{(b)}
 \nn \\
\Omega_{h_c,1}^{(b)} &=& - \Omega_{\chi_{c1},1}^{(b)} + ( w - 1 ) \Omega_{\chi_{c_1}, 2}^{(b)}
\\
\Omega_{h_c,2}^{(b)} &=& \Omega_{\chi_{c1},2}^{(b)} \nn
\eea

\bea
\Omega_{h_c,1}^{(c)} &=& \frac{1}{3} \Omega_{\chi_{c0}}^{(c)} + 2 \Omega_{\chi_{c1},1}^{(c)} + ( w - 1 ) \Omega_{\chi_{c1},2}^{(c)} + \frac{w - 1}{3} \big[ 3 \Omega_{\chi_{c2},1}^{(c)} + 2 \Omega_{\chi_{c2},2}^{(c)} \big]
\nn \\
\Omega_{h_c,2}^{(c)} &=& - \frac{1}{2} \Omega_{\chi_{c1},2}^{(c)} - \frac{1}{2} \big[ 3 \Omega_{\chi_{c2},1}^{(c)} + 2 \Omega_{\chi_{c2},2}^{(c)} \big]
\eea

\bea
\Omega_{h_c,1}^{(Q),\text{mix}} &=& - \frac{1}{3} \Omega_{\chi_{c0}}^{(Q),\text{mix}} - 2 \Omega_{\chi_{c1},1}^{(Q),\text{mix}} + (w+1) \Omega_{\chi_{c1},2}^{(Q),\text{mix}} - \frac{w + 1}{3} \big[ 3 \Omega_{\chi_{c2},1}^{(Q),\text{mix}} - 2 \Omega_{\chi_{c2},2}^{(Q),\text{mix}} \big]
\nn \\
\Omega_{h_c,2}^{(Q),\text{mix}} &=& \frac{1}{2} \Omega_{\chi_{c1},2}^{(Q),\text{mix}} - \frac{1}{2} \big[ 3\Omega_{\chi_{c2},1}^{(Q),\text{mix}} - 2 \Omega_{\chi_{c2},2}^{(Q),\text{mix}} \big] \, .
\eea

\bibliographystyle{JHEP}
\bibliography{refFFMNP1}
\end{document}